\documentclass[%
 aip,
 amsmath,amssymb,
 reprint,%
]{revtex4-1}

\usepackage{scalerel,tikz}
\usetikzlibrary{svg.path}
\definecolor{orcidlogocol}{HTML}{A6CE39}
\tikzset{orcidlogo/.pic={\fill[orcidlogocol] svg{M256,128c0,70.7-57.3,128-128,128C57.3,256,0,198.7,0,128C0,57.3,57.3,0,128,0C198.7,0,256,57.3,256,128z}; \fill[white] svg{M86.3,186.2H70.9V79.1h15.4v48.4V186.2z} svg{M108.9,79.1h41.6c39.6,0,57,28.3,57,53.6c0,27.5-21.5,53.6-56.8,53.6h-41.8V79.1z M124.3,172.4h24.5c34.9,0,42.9-26.5,42.9-39.7c0-21.5-13.7-39.7-43.7-39.7h-23.7V172.4z} svg{M88.7,56.8c0,5.5-4.5,10.1-10.1,10.1c-5.6,0-10.1-4.6-10.1-10.1c0-5.6,4.5-10.1,10.1-10.1C84.2,46.7,88.7,51.3,88.7,56.8z};}}
\newcommand\orcidicon[1]{\href{https://orcid.org/#1}{\mbox{\scalerel*{
\begin{tikzpicture}[yscale=-1,transform shape]\pic{orcidlogo};
\end{tikzpicture}}{|}}}}
\usepackage{graphicx}
\usepackage{dcolumn}
\usepackage{bm}
\usepackage{hyperref}
\usepackage[utf8]{inputenc}
\usepackage[T1]{fontenc}
\usepackage{mathptmx}
\usepackage{etoolbox}
\usepackage{caption}
\usepackage{subcaption}
\usepackage{tikz}
\hypersetup{
  colorlinks   = true, 
  urlcolor     = blue, 
  linkcolor    = blue, 
  citecolor   = blue 
}
\makeatletter
\def\@email#1#2{%
 \endgroup
 \patchcmd{\titleblock@produce}
  {\frontmatter@RRAPformat}
  {\frontmatter@RRAPformat{\produce@RRAP{*#1\href{mailto:#2}{#2}}}\frontmatter@RRAPformat}
  {}{}
}%
\makeatother
\begin{document}

\preprint{AIP/123-QED}

\title[]{Spatiotemporal dynamics of transonic shock-wave/turbulent-boundary-layer interactions in an overexpanded planar nozzle}
\author{Justin Kin Jun Hew}
 \email{u7322062@anu.edu.au}
\affiliation{Space Plasma Power and Propulsion (SP3) Laboratory, Research School of Physics, Australian National University, Canberra, ACT 2601, Australia}%
\affiliation{Research School of Astronomy and Astrophysics, Australian National University, Canberra, ACT 2611, Australia}

\author{Emanuele Martelli}
\affiliation{Department of Engineering, University of Campania “L. Vanvitelli”, Via Roma 29, 81031, Aversa, Italy}

\author{Mahdi Davoodianidalik}
\affiliation{%
Space Plasma Power and Propulsion (SP3) Laboratory, Research School of Physics, Australian National University, Canberra, ACT 2601, Australia
}%
\affiliation{%
Physics of Fluids Laboratory, Research School of Physics, Australian National University, Canberra, ACT 2601, Australia
}%
\author{Rod W. Boswell}
\affiliation{%
Space Plasma Power and Propulsion (SP3) Laboratory, Research School of Physics, Australian National University, Canberra, ACT 2601, Australia
}%
\author{Christoph Federrath}
\affiliation{Research School of Astronomy and Astrophysics, Australian National University, Canberra, ACT 2611, Australia}%
\affiliation{Australian Research Council Centre of Excellence in All Sky Astrophysics (ASTRO3D), Canberra, ACT 2611, Australia}
\author{Matthew Shadwell}
\affiliation{Space Plasma Power and Propulsion (SP3) Laboratory, Research School of Physics, Australian National University, Canberra, ACT 2601, Australia}%

\date{\today}

\begin{abstract}
We perform a combined numerical and experimental study to investigate the transonic shock-wave/turbulent-boundary-layer interactions (STBLI) in a shock-induced separated subscale planar nozzle with fully-expanded Mach number, $M_j = 1.05$ and jet Reynolds number $Re \sim 10^{5}$. The nozzle configuration is tested via time-resolved schlieren visualisation.  While numerous studies have been conducted on the high Reynolds number separated flowfields, little is known on the weak shock wave unsteadiness present in low nozzle pressure ratio (NPR) transonic nozzles. Therefore, numerical simulations are carried out with high resolution three-dimensional delayed detached eddy simulation (DDES), to study the spatiotemporal dynamics of wall pressure signals and unsteady shock interactions.  The transient statistics considered include spectral Fourier and wavelet-based analysis and dynamic mode decomposition (DMD). The spectral analyses reveal energetic low frequency modes corresponding to the staging behaviour of shock unsteadiness, and high frequencies linked to the characteristics of the Kelvin-Helmholtz instabilities in the downstream turbulent mixing layer. The mechanisms for the low frequency unsteadiness is educed through modal decomposition and spectral analysis, wherein it is found that the downstream perturbations within the separation bubble play a major role in not only closing the aeroacoustic feedback loop, but allowing the continual evolution and sustainment of low frequency unsteadiness. An analysis via the vortex sheet method is also carried out to characterise the screech production, by assuming an upstream propagating guided jet mode. 
\end{abstract}

\maketitle

\section{Introduction}\label{sec:level1}

The study of shock-wave turbulent boundary layer interactions (STBLI) occurring in supersonic internal and external flows has received a significant amount of interest, due to its great importance in many engineering applications of interest. These include the design of supersonic inlets \citep{holden2005separated,babinsky2008sbli,ogawa2010inlet},  impinging jets \citep{henderson2005experimental}, rocket nozzles \citep{martelli2017detached,della2019enhanced}, scramjet engines, biconic bodies and others (see e.g., review articles by Dolling  \cite{dolling2001fifty}, Hadjadj and Onofri \cite{hadjadj2009nozzle} and Gaitonde \cite{gaitonde2015progress}). Transonic and supersonic propulsive nozzles operating at off-design conditions often result in shock-induced separation, which induces dynamical instabilities, wall-pressure oscillations, generation of off-axis forces, aeroacoustic resonance such as screech, transonic tones and Mach wave radiation\cite{deck2012recent,martelli2017detached,martelli2019characterization,martelli2020flow,bakulu2021jet}. These effects can even result in engine unstart and structural damage to the engine in question \citep{nave1973sea}.  

It has been identified that there are three main frequency components in unsteady shock boundary layer interaction, as observed in canonical configurations of oblique shock SBLI (OSWBLI) (swept fins, double cones, impinging SBLI etc.) (see e.g.~\citet{clemens2014low} for a review), which all scale with the freestream velocity $U_{\infty}$, and the $99 \%$ boundary layer thickness, $\delta_{99}$. High frequency unsteadiness is of order $f \sim \mathcal{O}(U_{\infty}/ \delta_{99})$,  which is triggered by the incoming turbulent boundary layer (TBL) as a result of the evolution of coherent structures and anisotropic nature of the upstream flow. The intermediate range, which is of order $f \sim \mathcal{O}(0.1U_{\infty}/\delta_{99})$, characterises the vortex shedding resonant frequency from the separated shear layer, which is dominated by oblique modes, displaying fundamental characteristics that is similar to the incompressible mixing layer. The low frequency unsteadiness, which is of order $f \sim \mathcal{O}(0.01U_{\infty}/\delta_{99})$, is related to the unsteadiness located at the shock-induced point of separation. There are two dominant theories about the origin of this inherent low frequency dynamics, one of which suggests that the incoming boundary layer carries inherent low order unsteadiness through turbulent fluctuations, which promotes and modulates the spectral signature at the separation point \citep{plotkin1975shock,beresh2002relationship,ganapathisubramani2006large,ganapathisubramani2007effects}. Such a hypothesis was first observed in experiments by \citet{andreopoulos1987some} for the compression-ramp flows (CRSBLI) at freestream Mach number $M_{\infty} = 2.84 $, and has been observed numerically and analytically by \citet{touber2011low} using large eddy simulations (LES) and stochastic (Fokker-Planck) analytical modelling approaches, where low frequency global modes were identified and related to superstructures in the upstream TBL. 

The second theory relates the unsteadiness frequencies to downstream propagating disturbances within the separation bubble possibly resulting from shock-expansion bubble interactions \citep{pirozzoli2006direct,dupont2006space,touber2009large}. Since the separation bubble is elliptic in nature (subsonic flow), acoustic waves propagating downstream can thus re-transmit upstream to close the feedback loop and support the low frequency content. These flow behaviours are made possible by the high adverse pressure gradient imposed on the mean flow during flow separation \citep{na1998structure}, where intermittent expansion and contraction of the entrainment region within the separation bubble supports continuous upstream and downstream travelling waves \citep{piponniau2009simple}. 

As demonstrated above, substantial work has been conducted on the behaviour of SBLI in canonical configurations. With regards to off-design supersonic nozzles exhibiting shock-induced flow separation, similar features occur by analogy, where wall-pressure perturbations at the separation shock location can induce local flow unsteadiness \citep{deck2002numerical,ostlund2002flow}, and subsequent interactions between the incident shock and separation bubble can lead to severe side load generation \citep{chen1994numerical,schmucker1984flow,dumnov1996unsteady,xiao2007numerical} (see e.g., Fig.1 of~\citet{bakulu2021jet}). 

In these cases, the source of the low frequency unsteadiness and mechanisms of its production, has been a subject of extensive numerical and experimental studies\citep{verma2014effect,jaunet2017wall,verma2018analysis,martelli2017detached,martelli2019characterization,martelli2020flow,zebiri2020shock,bakulu2021jet}; where most of the work on boundary layer separation has been concentrated on the high area ratio axisymmetric jets, such as the Thrust Optimised Contour (TOC) and TIC (Truncated Ideal Contour) \citep{hagemann2008shock,baars2012wall,baars2013transient} nozzles, or the more recent dual-bell nozzle (e.g., \citet{martelli2007numerical,cimini2021numerical}), due to its special relevance with the well-known engines of today. These exhibit a range of different types of shock structures, depending on the nozzle pressure ratio (NPR), such as the Free Shock Separation (FSS) and Restricted Shock Separation (RSS). However, here we only consider planar jets, which only exhibit the FSS \citep{hadjadj2009nozzle}, which have interestingly only been studied sparsely throughout the years \cite{xiao2007numerical,papamoschou2009supersonic,johnson2010instability,olson2011large,verma2014origin,zebiri2020shock}.  

A crucial part of the enhanced wall-pressure perturbations through the emergence of a separation shock structure is its consequent generation of aeroacoustical resonance. One of these prominent discrete tones recorded in low area ratio nozzles is the so-called transonic resonance \citep{zaman2002investigation}, associated with a feedback loop between upstream and downstream propagating waves from the shear layer that is approximately equal to frequencies one quarter of an acoustic standing wave in an open pipe of length L \citep{martelli2020flow,zebiri2020shock}. Another type of sound generation, through possibly broadband-shock associated noise (BBSAN) or screech \citep{tam1986proposed}, follows a similar pathway where a feedback loop is always maintained between waves originating from the initial separation zone, to the downstream acoustic waves that radiate back upstream \citep{olson2013mechanism}.

Moreover, we note that \citet{larusson2017dynamic} has also shown that just by conducting Dynamic Mode Decomposition (DMD) with two-dimensional Reynolds-averaged Navier-Stokes (RANS), they could already identify the transonic resonant tone observed by \citet{zaman2002investigation}, where the frequencies and flow structures are consistent with that of a hydrodynamically supported standing acoustic wave. 

Nevertheless, the problem of fluid-structure interactions within nozzles undergoing flow separation still remains an active area of research today, hence the objective of the current study. 

Thus, we here conduct a combined experimental and numerical study using Delayed Detached Eddy Simulation (DDES) on a subscale over-expanded transonic planar nozzle of throat height $2$ mm, with fully-expanded Mach number $M_j = 1.05$. 

Such small subscale nozzle configurations have importance in not only understanding the salient characteristics of large transonic and supersonic rocket nozzles as a whole, but also has major implications for miniature propulsion technologies, such as Cubesats, or low-thrust rockets, possibly utilising microjets (see e.g.~\citet{xu2007two,lijo2015analysis,nazari2020multi}), which are still not thoroughly understood on a fundamental level (e.g., reviews by \citet{louisos2008design} and \citet{levchenko2018space}), thus warranting further studies especially considering the effect of low operating Reynolds number, to the laminar-to-turbulent boundary layer transition during separation.

The rest of the paper is organised as follows. In Sec.~\ref{sec2:experiment} and Sec.~\ref{sec3:numerical} we discuss our experimental and numerical methodologies, respectively. Then in Sec.~\ref{sec4:results} we present the analysis of our numerical results, including spectral, wavelet and modal decomposition techniques, with comparison to some experimental flow visualisations. Finally, Sec.~\ref{sec:conclusion} summarises the main results and conclusions of the study.

\section{Experimental method}\label{sec2:experiment}

The sub-scale nozzle used in this study was designed via high-precision CNC machining with a tight tolerance of $\pm 10 \mu m$. It is a minimum length supersonic planar nozzle of design Mach number, $M_d = 3 $, which has been constructed with the use of a method of characteristics (MoC) code, creating a nozzle throat-to-exit aspect ratio of $A_r = 4.2$, with throat height $h_t = 2$mm. The minimum length nozzle follows a standard truncation that excludes the expansion section of the divergent contour, leaving only the expansion section. The spanwise direction extends $z = 5h_t$, where $h_t = 2$mm is the nozzle throat height. In the test setup, the nozzle operates in blow-down mode, with a nozzle inlet pressure of $P_i = 2 \times 10^5$ Pa exiting into ambient, yielding an $NPR$ of about 2 based on measurements via a pressure transducer. Two BK-7 optical glass slabs are attached to the sides of the diverging section of the nozzle, allowing for visualisation of the internal flow.  

The small-scale nature of the jet, especially considering the low NPR nature of the flow, thus results in rather weak shock formation. Such interactions thus require high resolution methods for adequate visualisation of the internal flow, as described below. 

Flow visualisation is conducted at the Supersonic Wind Tunnel laboratory at the University of New South Wales, Canberra (ADFA). A standard Z-type monochrome schlieren with a straight knife edge is used for time-resolved visualisation of the internal shock structure within the separated nozzle flow. Images are taken with a high-speed camera (Shimadzu HPV-X2) at 100,000 frames per second with an individual frame exposure time of 10 microseconds. The spatial resolution is 0.3 mm/pixel. With the given spatial and temporal resolution, it is difficult to resolve flow features with sub-millimetre dimensions (e.g., Mach stems\citep{kleine2014influence}). 

More details of the experimental configuration and the test facility, along with details of similar operations with the shadowgraph method, can be found in \citet{kleine2014influence} and \citet{skews2007flow,skews2010shock}.
 \begin{figure}
    \centering
     \begin{subfigure}[b]{0.46\textwidth}
         \centering
         \includegraphics[trim = 2.0cm 4cm 2cm 3cm,width=\textwidth]{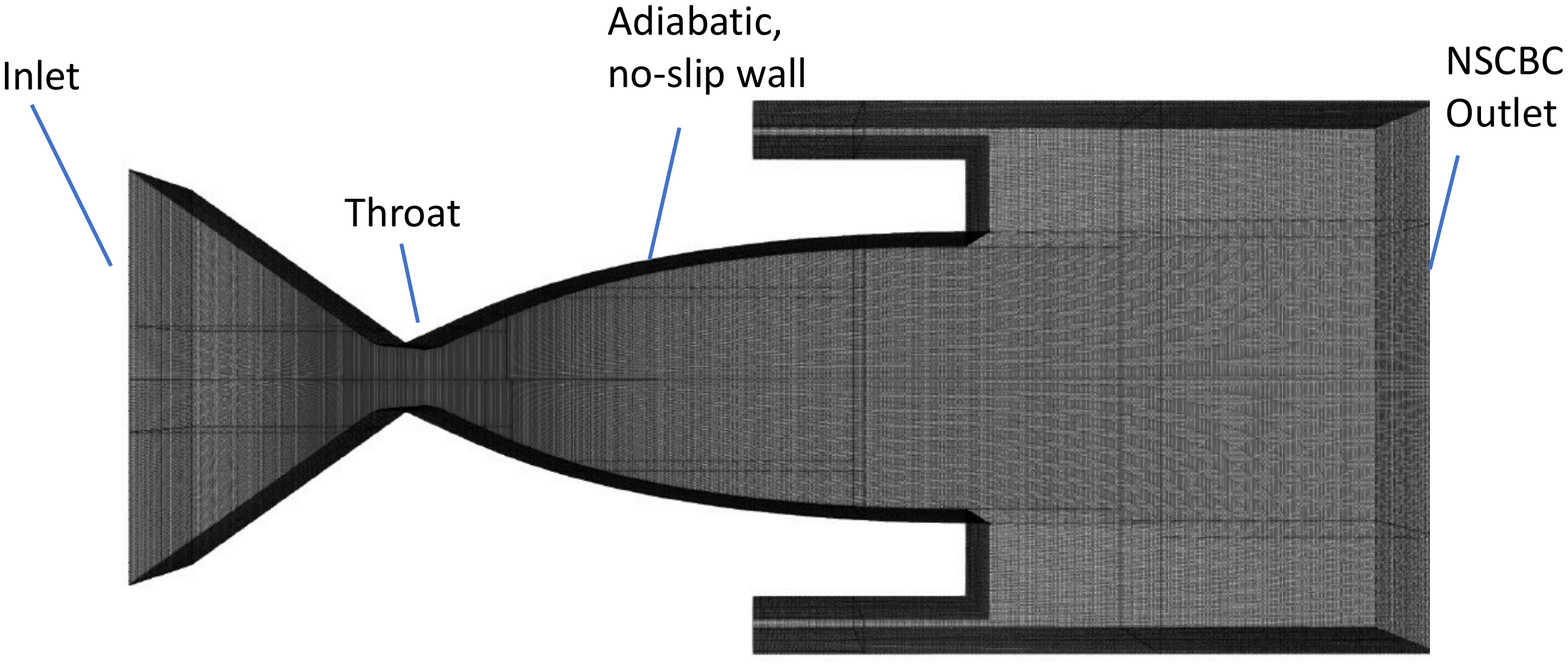}
         \caption{\label{fig:meshx}}
         
     \end{subfigure}
     \vspace{0.05cm}
     \begin{subfigure}[b]{0.46\textwidth}
         \centering
         \includegraphics[trim = 2.0cm 2cm 2cm 1cm,width=\textwidth]{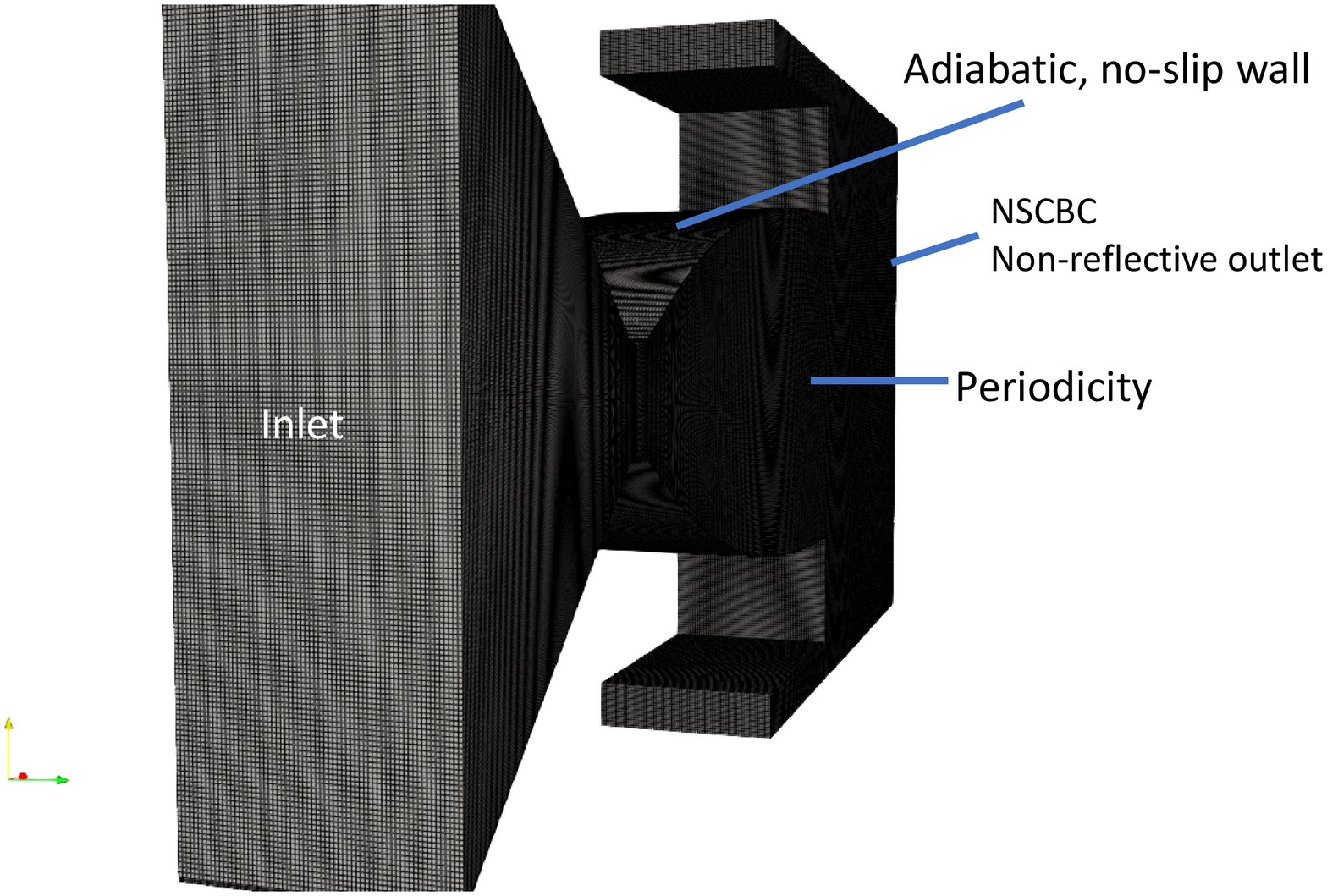}
         \caption{\label{fig:meshy}}
         
     \end{subfigure}
             \caption{\label{fig:mesh} Three-dimensional computational mesh topology of the minimum-length ideal contour nozzle used in the current study, with the annotated main locations of interest (i.e., inlet, wall, throat, outlet) and boundary conditions (e.g., no-slip, periodicity and Navier-Stokes Characteristic Boundary Condition), (a) 2D view, (b) lateral 3D view. }
        
\end{figure}
\section{Numerical simulations}\label{sec3:numerical}
We conduct delayed detached eddy simulations (DDES) \citep{spalart2006new}, which is a hybrid RANS/LES method that treats the upstream flow in the attached boundary layer through the Reynolds-Averaged Navier Stokes (RANS) method, while the separated core flow is treated in the large eddy simulation (LES) mode.  This method addresses the problem of modelled stress depletation previously observed in the traditional DES approaches, while still providing greater resolution of massively separated flows through the LES framework.
\subsection{Physical Modelling}
We solve the three-dimensional Reynolds averaged/Favre-filtered Navier-Stokes equations for a compressible, calorically perfect gas with viscous heating:
\begin{equation}
\frac{\partial \rho}{\partial t}+\frac{\partial\left(\rho u_j\right)}{\partial x_j}=0,
\end{equation}
\begin{equation}
\frac{\partial\left(\rho u_i\right)}{\partial t}+\frac{\partial\left(\rho u_i u_j\right)}{\partial x_j}+\frac{\partial p}{\partial x_i}-\frac{\partial \tau_{i j}}{\partial x_j}=0,
\end{equation}
\begin{equation}
\frac{\partial(\rho E)}{\partial t}+\frac{\partial\left(\rho E u_j+p u_j\right)}{\partial x_j}-\frac{\partial\left(\tau_{i j} u_i-q_j\right)}{\partial x_j}=0,
\end{equation}
where $\rho$, $u_i$ ,$p$, $E$ and $q$ are gas density, velocity componeents, pressure, specific total energy (thermal and kinetic) respectively.  The use of either the standard Reynolds-averaged or a Favre-filtered form depends on whether the RANS or LES branch is active. The total stress tensor $\tau_{ij}$ is the sum of the viscous and Reynolds stress:
\begin{equation}
    \tau_{i j}=2 \rho\left(v+v_t\right) S_{i j}^* \quad S_{i j}^*=S_{i j}-\frac{1}{3} S_{k k} \delta_{i j},
\end{equation}
where $S^*_{ij}$ is the traceless rate of strain tensor, which is the symmetric part of the velocity gradient tensor $\partial_i u_j$ that accounts for physical shear viscosity. Here, we have used the fact that the eddy-viscosity hypothesis applies, by introducing the turbulent viscosity $\nu_{t}$, which is to be modelled with a subgrid scale model. Sutherland's model calculates the temperature-dependent kinematic viscosity, and the total heat $q_j$ is modelled using Newton's law for heat conduction.
\subsection{Turbulence model}
We use the Spalart-Allmaras variant of the DDES formulation, where instead of solving the eddy viscosity directly, we solve for the Spalart-Allmaras variable $\tilde{\nu}$:
\begin{equation}
\begin{aligned}
\frac{\partial(\rho \tilde{v})}{\partial t}+\frac{\partial\left(\rho \tilde{v} u_j\right)}{\partial x_j}=c_{b 1} \tilde{S} \rho \tilde{v}+\frac{1}{\sigma}\left[\frac{\partial}{\partial x_j}\biggl[(\rho v+\rho \tilde{v}) \frac{\partial \tilde{v}}{\partial x_j}\right] \\+c_{b 2} \rho\left(\frac{\partial \tilde{v}}{\partial x_j}\right)^2\biggl]-c_{w 1} f_w \rho\left(\frac{\tilde{v}}{\tilde{d}}\right)^2
\end{aligned}
\end{equation}
where $\tilde{S}$ is the modified vorticity magnitude, $f_{\omega}$ is the near-wall damping function, $c_{b1}$, $c_{b2}$, $c_{\omega1}$ and $\sigma$ are model constants from \citet{spalart2006new}. The eddy viscosity is then solved with a correction function, $\nu_t = f_{v1}\tilde{\nu}$, which guarantees the correct boundary layer flow profile in the near-wall region. We also define a modified length scale, $\tilde{d}$ as:
\begin{equation}
    \tilde{d}=d_w-f_d \max \left(0, d_w-C_{D E S} \Delta\right)
\end{equation}
to determine the transition between RANS and LES mode, ensuring that the attached boundary layer is always treated in RANS, even with ambiguous grid densities in the so-called ``grey area". Here $C_{\textrm{DES}}$ is a calibration constant, which we set to 0.20, and $f_d$ is defined as:
\begin{equation}
    f_d=1-\tanh \left[\left(16 r_d\right)^3\right], \quad r_d=\frac{\tilde{v}}{k^2 d_w^2 \sqrt{U_{i, j} U_{i, j}}}
\end{equation}
where $k$ is the von Kármán constant and $U_{i,j}$ is the velocity gradient tensor. Note that these constants differ from the original DDES, and are based on calibration tests done by \citet{martelli2017detached,martelli2019characterization,martelli2020flow}. $f_d$ acts as a shielding function that enforces RANS treatment at the wall even if one performs a wall-resolved simulation. Moreover, here our subgrid length scale $\Delta$ is dependent on the flowfield itself: 
\begin{equation}
    \Delta=\frac{1}{2}\left[\left(1+\frac{f_d-f_{d 0}}{\left|f_d-f_{d 0}\right|}\right) \Delta_{\max }+\left(1-\frac{f_d-f_{d 0}}{\left|f_d-f_{d 0}\right|}\right) \Delta_{v o l}\right]
\end{equation}
based on the method proposed by \citet{deck2012recent}, where $f_{d0} = 0.9 $ and $\Delta_{\textrm{max}} =\mathrm{max}(\Delta x, \Delta y, \Delta z) $ and $\Delta_{\mathrm{vol}} = (\Delta x \cdot \Delta y \cdot \Delta z)^{1/3}$. Ultimately, the function $f_d$ ensures that the subgrid length scale depends intrinsically on the flow, with the addition of the second term on the right hand side that suppresses subgrid scale viscosity, preventing the restriction of coherent structures in the shear layer. This method is different from the original DDES, and has been implemented based on extensive compressible turbulent boundary layer (TBL) studies (see \citet{deck2020towards}).

\begin{figure*}
     \centering
     \begin{subfigure}[b]{0.46\textwidth}
         \centering
         \includegraphics[width=\textwidth]{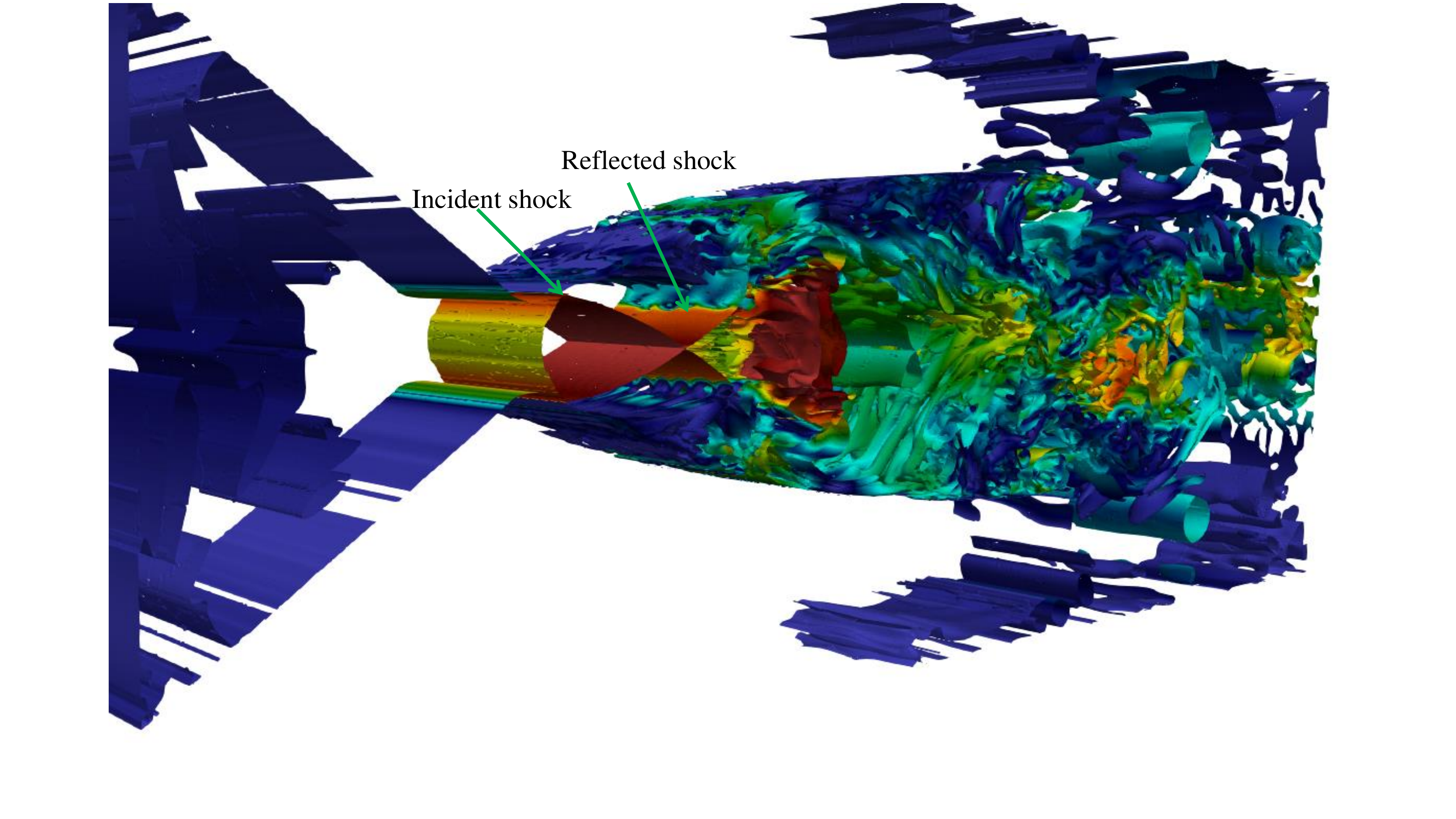}
         \caption{\label{fig:Bxprime}}
         
     \end{subfigure}
     \vspace{0.05cm}
     \begin{subfigure}[b]{0.46\textwidth}
         \centering
         \includegraphics[trim= 1cm 3cm 1cm 1cm,width=\textwidth]{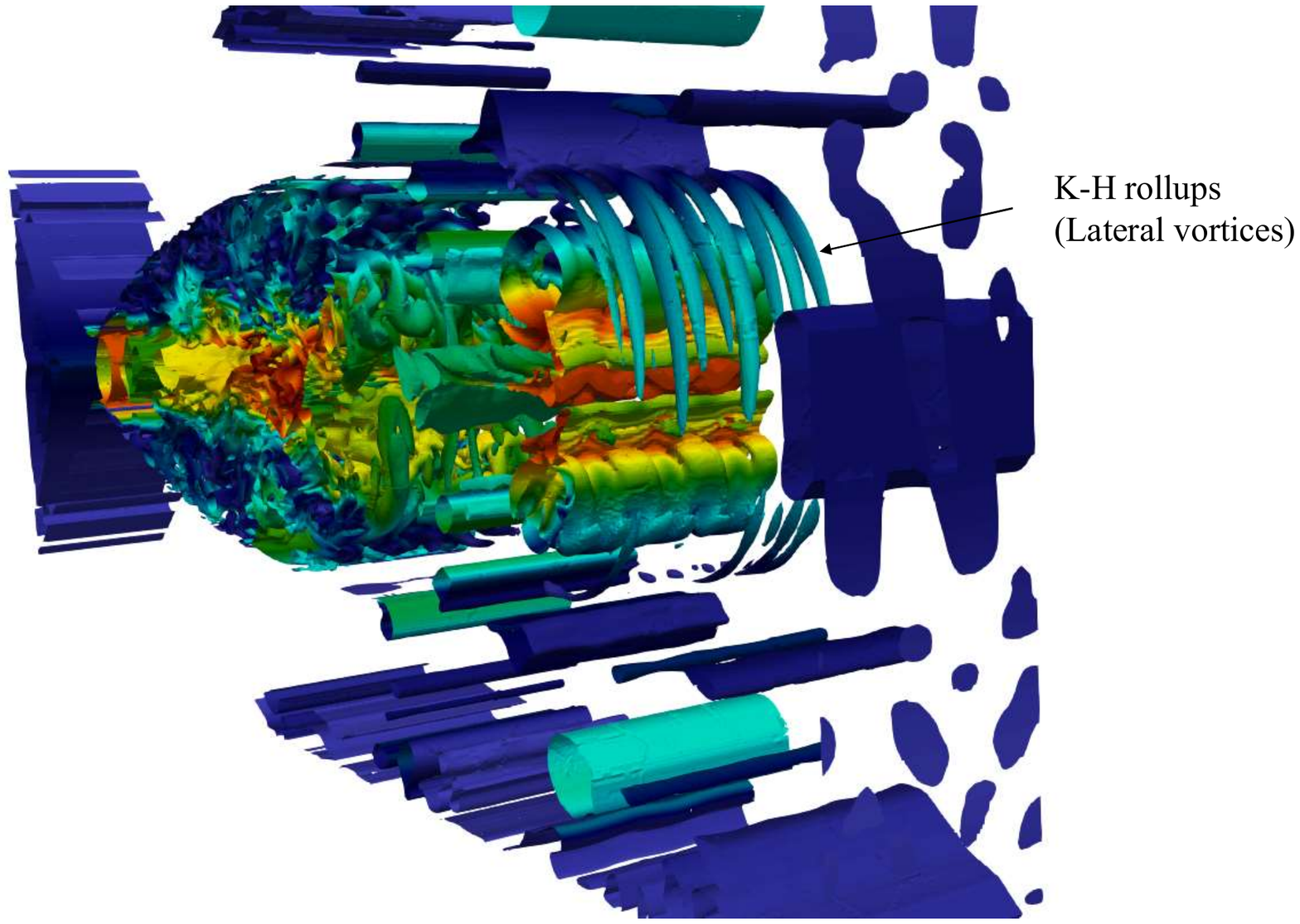}
         \caption{\label{fig:Byprime}}
         
     \end{subfigure}
        \caption{Iso-volume of the three-dimensional Q-criterion, overlaid with a numerical schlieren $\lvert \nabla \rho \rvert$, coloured by velocity magnitude $\lvert \mathbf{u} \rvert$. (a) Annotated are the locations of the shock-induced separation points, where the incident and reflected shocks form, along with a large scale separation bubble above. (b) Counter-rotating Kelvin-Helmholtz instabilities are visible, they originate from the turbulent shear layer induced through the passage of the shock. Enlarged versions available in Appendix~\ref{sec:qCriterion}. }
        \label{fig:qcriterions}
\end{figure*}

\subsection{Numerical Solver}
A structured compressible flow solver based on a modified OpenFOAM framework \citep{weller1998tensorial} is used in this study, which has been developed and modified based on the well-validated HiSA library \citep{heyns2014modelling}. Convective and diffusive fluxes are discretised using a mixed Total Variation Diminishing (TVD) second-order upwind scheme and third order van Leer method. The numerical flux functions are calculated using the improved AUSM+UP  flux vector splitting scheme with low-speed correction of \citet{liou2006sequel}, and the bounded Green-Gauss interpolation of the cell-centred values to face values are used to reconstruct the respective fluxes. 

We also use a dual-time stepping method with the explicit second order Euler backward differencing for time discretisation and advancement. This improves the stability of the numerical system substantially, where the solution is advanced both in a pseudo-time step, as well as in real time, exploiting the pseudo-transient methods employed commonly for steady-state flows. Finally, the matrix-free generalised minimal residual method (GMRES) of \citet{saad1986gmres} is used to solve the linearised system iteratively until convergence at each time step. 
\subsection{Geometry and initial conditions}
Fig.~\ref{fig:mesh} displays the computational domain used in the current study. Since the upstream flow is treated in RANS mode, we ensure that the near-wall grid satisfies $\Delta y^ + <1$ in order to resolve the laminar sublayer. We further ensure that all LES grid requirements in the streamwise and spanwise extent are satisfied \citep{spalart2006new}, with $\Delta z^+ <30$ and $\Delta x^+ <30$, yielding a grid density of about $14.3 \times 10^6$ cells. This grid density is also very similar to that used in prior planar DDES nozzle works \citep{deck2009delayed,martelli2017detached,martelli2019characterization}.
The Navier-Stokes Characteristics Boundary Conditions (NSCBC) method of \citet{poinsot1992boundary} is applied to ensure non-reflective inflow and outflow boundary conditions at the nozzle inlet and outlet, which uses Riemann invariants to ensure that no outgoing characteristics are reflected back into the domain. The flow is assumed to be periodic in the spanwise direction, where $z = 2h_t$, and all walls are no-slip and adiabatic. We note that the spanwise length simulated here does not correspond to that used in the experiment, and is truncated here in order to save computational time. We show, in the following section, that this plays little to no role in containing the development of spanwise coherent structures.

Here we also quantify the Reynolds number based on the stagnation chamber parameters, which is:
\begin{equation}
    \text{Re}  = \frac{\rho_0 a_0 h_t/2}{ \mu_0 } =\frac{\sqrt{\gamma}}{\mu(T)} \frac{p_0 h_t / 2}{\sqrt{R_{\mathrm{air}} T_0}} \approx 2.2 \times 10^5 
\end{equation}
where $p_0$ is the pressure in the stagnation chamber, $\mu(T)$ is the temperature-dependent dynamic viscosity based on Sutherland's law, $R_{air}$ is the specific gas constant and $\rho_0$, $a_0$ and $T_0$ are the blowdown density, acoustic speed and temperature respectively.
The relevant non-dimensional parameters such as the fully-expanded Mach number,~$M_j$, jet height $h_j$ and velocity, $u_j$ are  also computed according to the well-known relations by \citet{tam1988shock}. We also further evaluate the Reynolds number based on the $99 \% $ boundary layer thickness $\delta_{99}$, which  is $Re_{\delta_{99}} = 2.5 \times 10^4$.

The simulation ran for about 400000 time steps, with a maximum real time CFL of 0.8, for a total of already about 360 thousand core hours. In non-dimensional units, this corresponds to a physical time of $60 h_j/ U_j $, which is in between the ranges of prior works. \citet{nichols2011high} conducted simulations with a flow time $45 D_{eq} / U_j$ to characterise aeroacoustic resonance, \citet{gojon2019antisymmetric} used a simulation time of $500 D_{eq}/ U_j$, where $D_{eq}$ is the equivalent jet diameter, while \citet{bogey2012influence} used $75 D/U_j - 100 D/u_j$ to study screech generation in a fully-expanded subsonic jet.  Thus, the simulation times here lie within the intermediate range of prior simulation works, and is sufficient to characterise the key components of the aeroacoustic feedback mechanisms, where the lowest resolvable Strouhal number, is as low as $St_{\text{min}} = f h_j / U_j \approx 0.015$. We further note that this resolves the lowest frequency tones observed in the spectral analysis, for at least a few cycles. While longer simulation times may be desirable, we demonstrate in the later sections that the low-frequency shock oscillations are still sufficiently well-captured for our purpose. 
\section{Results and Discussion}\label{sec4:results}
\subsection{Global flow features}
\begin{figure*}
     \centering
     \begin{subfigure}[b]{\columnwidth}
         \centering
         \includegraphics[width=\textwidth]{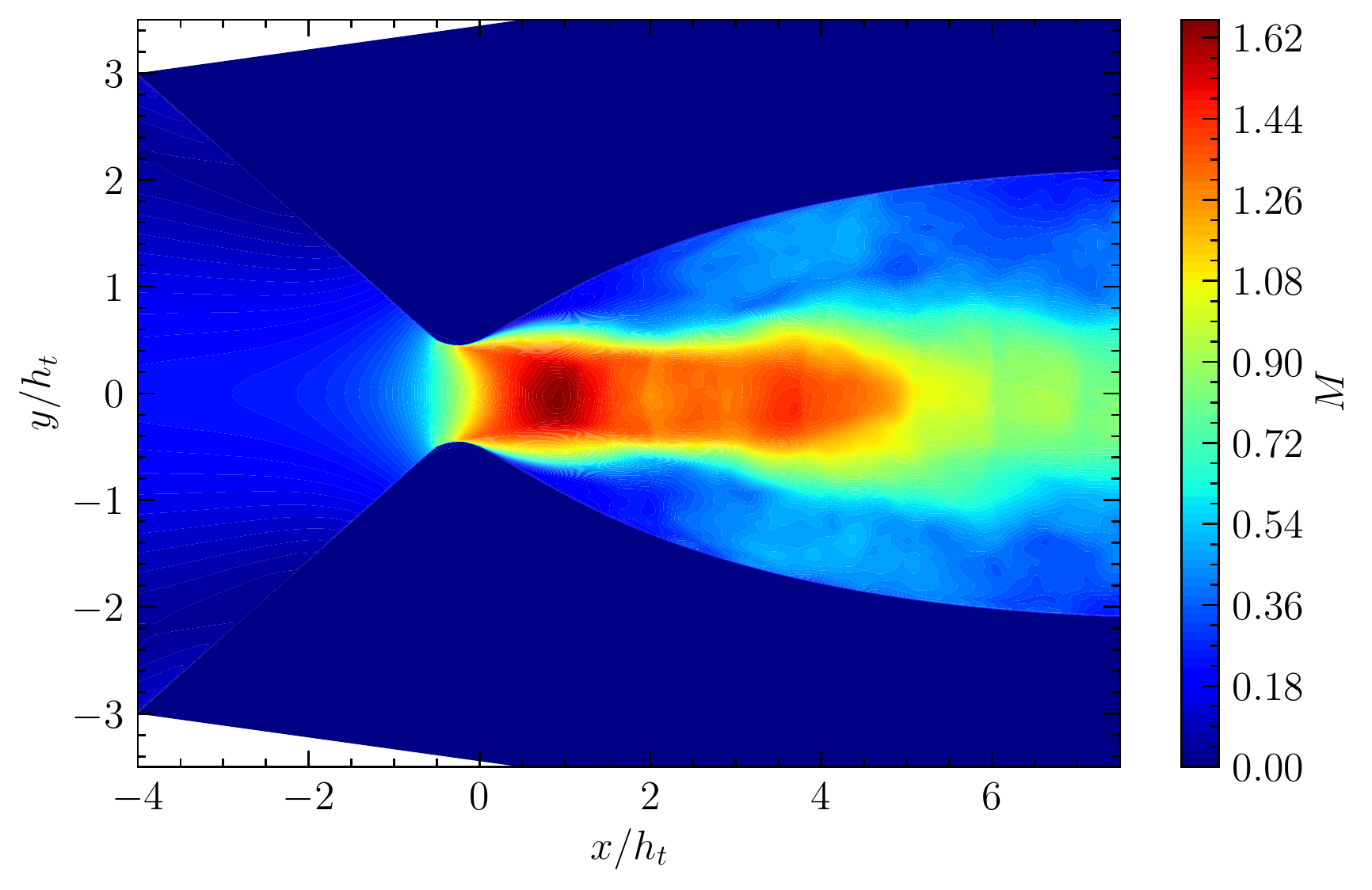}
         \caption{\label{fig:machcontour}}
     \end{subfigure}
     \begin{subfigure}[b]{\columnwidth}
         \centering
         \includegraphics[width=\textwidth]{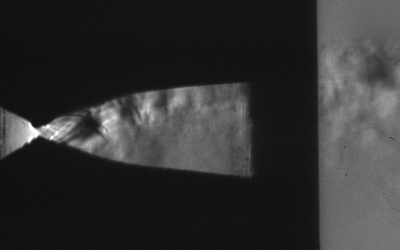}
         \caption{\label{fig:schlierne}}
         
     \end{subfigure}
        \caption{(a) Time-averaged Mach number contour, the field is time-averaged over 500 snapshots after the initial numerical transient. The fully-expanded Mach number, $M_j \approx 1.05$ at the separation point. (b) Schlieren image of the shock structure, with $\partial_{x} \rho$ density gradients at $NPR~\approx 2$ (note that this test in particular was not run with high resolution). A schlieren movie has been made available in the Supplementary Material. }
        \label{fig:compare}
\end{figure*}

Fig.~\ref{fig:qcriterions} displays the iso-volume of the computed Q-criterion, which is the second invariant of the velocity gradient tensor $\partial_{i} u_j$:
\begin{equation}
    Q = \frac{1}{2} (\lvert \Omega \rvert^2 - \lvert \mathcal{S} \rvert ^2 )
\end{equation}
where $\Omega$ and $\mathcal{S}$ are the anti-symmetric and symmetric parts of $\partial_{i} u_j$, respectively, called the rate of rotation tensor and strain rate tensor. It can be seen that large vortical structures dominate the flow past the incipient separation point, which is linked to the adverse pressure gradient imposed on the boundary layer profile \citep{na1998structure}. Moreover, the advected vortices and coherent structures are always in the wall-normal or spanwise directions (Kelvin-Helmholtz instabilities) \citep{touber2009large,martelli2020flow}, we do not find evidence of any streamwise Görtler vortices despite its possibility in the case of highly curved jet boundaries \citep{silnikov2014two} and canonical SWBLI flows \citep{grilli2013large,hu2021low}. Moreover, the shock structure is clearly positioned very near the nozzle throat; this is because the operating pressure conditions are especially low, so that the flow is transonic at separation $M_j = 1.05$, as shown also in the time-averaged Mach number contour in Fig.~\ref{fig:machcontour}. Thus, the shock structure is necessarily very weak, and only a regular reflection structure occurs with a two-shock configuration. 

Once the flow in the experiment was started, we observed substantial asymmetry in the shock generation even when the flow became steady. Such asymmetries were observed numerically only during the numerical transient, and were largely removed once a statistical time average was done. The exact reason for such flow asymmetries observed in low NPR nozzle experiments is ultimately still an open question \citep{papamoschou2009supersonic,johnson2010instability,verma2014origin}, and has been attributed to the so called ``Coanda effect" \citep{papamoschou2009supersonic,johnson2010instability,verma2018analysis}, where a wall-bounded detached jet always tends to reattach itself back to its surface. Parts of this could also be due to asymmetry in the nozzle geometry, which becomes relatively common in small-scale designs nearing the microscales owing to challenges with high-precision machining \citep{huang2007flow}. 

Nonetheless, Fig.~\ref{fig:compare} compares the time-averaged Mach number contour with the experimental schlieren image at a random time interval when the flow became steady. Reasonable agreement is found between the two in terms of the prediction of the separation point being very near the throat, even at such small nozzle length scales. The accuracy in the prediction of the separation point, which is the most important element that drives shock-induced unsteadiness and shock wave boundary layer interactions, thus validates the current simulation. We therefore continue our analysis with confidence of the fidelity of the numerical experiment.

\begin{figure}
\includegraphics[width = \columnwidth]{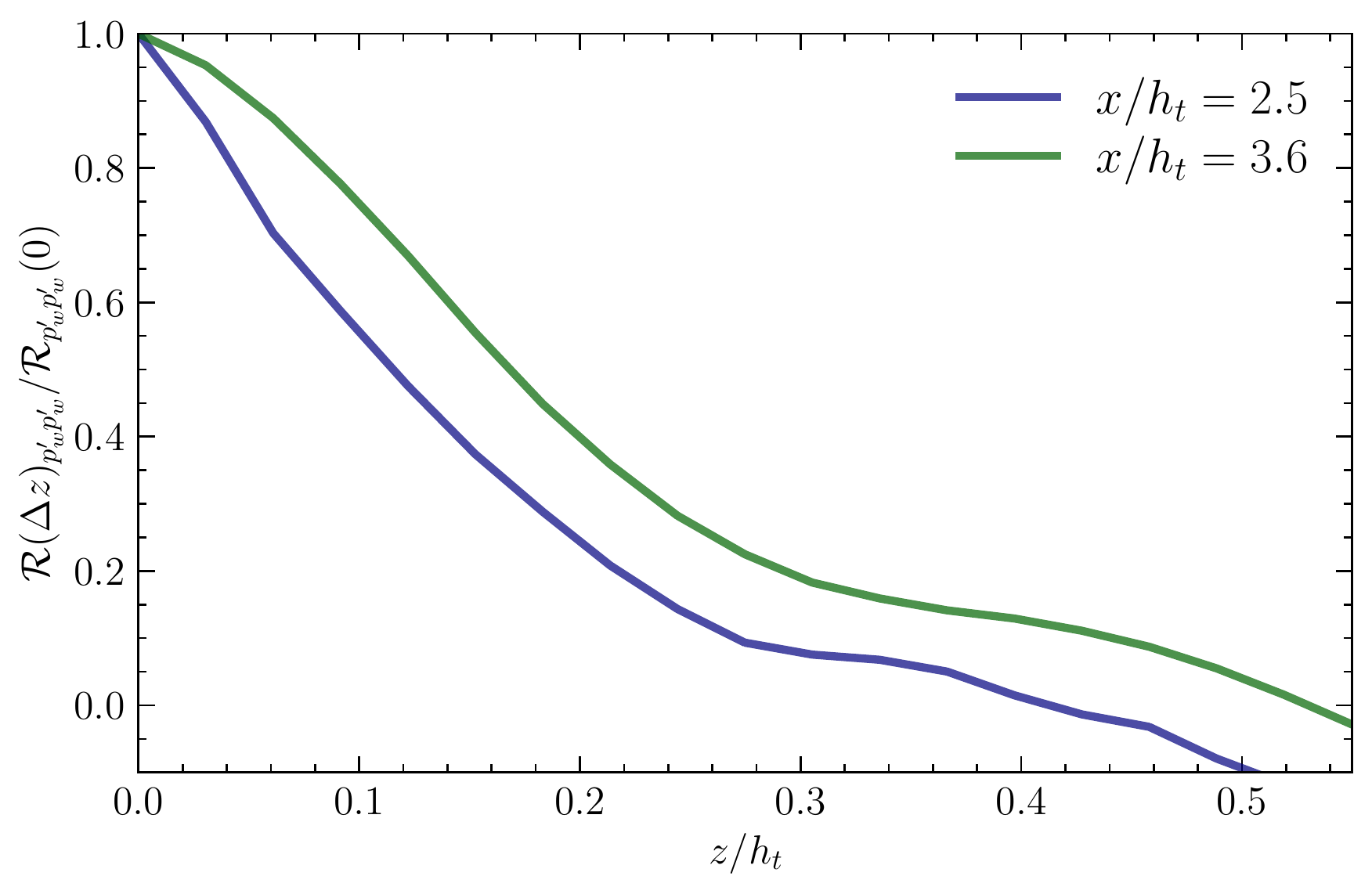}
\caption{\label{fig:twopointcorr} Normalised two-point spatial correlation function of the time-averaged wall-pressure fluctuations in the spanwise direction ($z$), across two different $x$-stations ($x/h_t = 2.5$ and $x/h_t=3.6$). The steep de-correlation indicates that spanwise coherent structures are not suppressed, and the spatial extent of $z = 2h_t$ is sufficient to resolve most of the energetic modes near the integral length scale.  }
\end{figure}

As a further step of validation, we plot the normalised two point correlation function in the spanwise direction, averaged across all time snapshots (Fig.~\ref{fig:twopointcorr}), the plot indicates that at $z= 0.5h_t$, a steep decorrelation occurs where all scales are no longer correlated with each other, indicating substantial anisotropy and turbulence production. This implies the integral length scale in the $z$ direction is not contained or suppressed.

\subsection{Turbulence amplification}

\begin{figure*}
     \centering
     \begin{subfigure}[b]{\columnwidth}
         \centering
         \includegraphics[trim = 1cm 3cm 1cm 3cm,width=\textwidth]{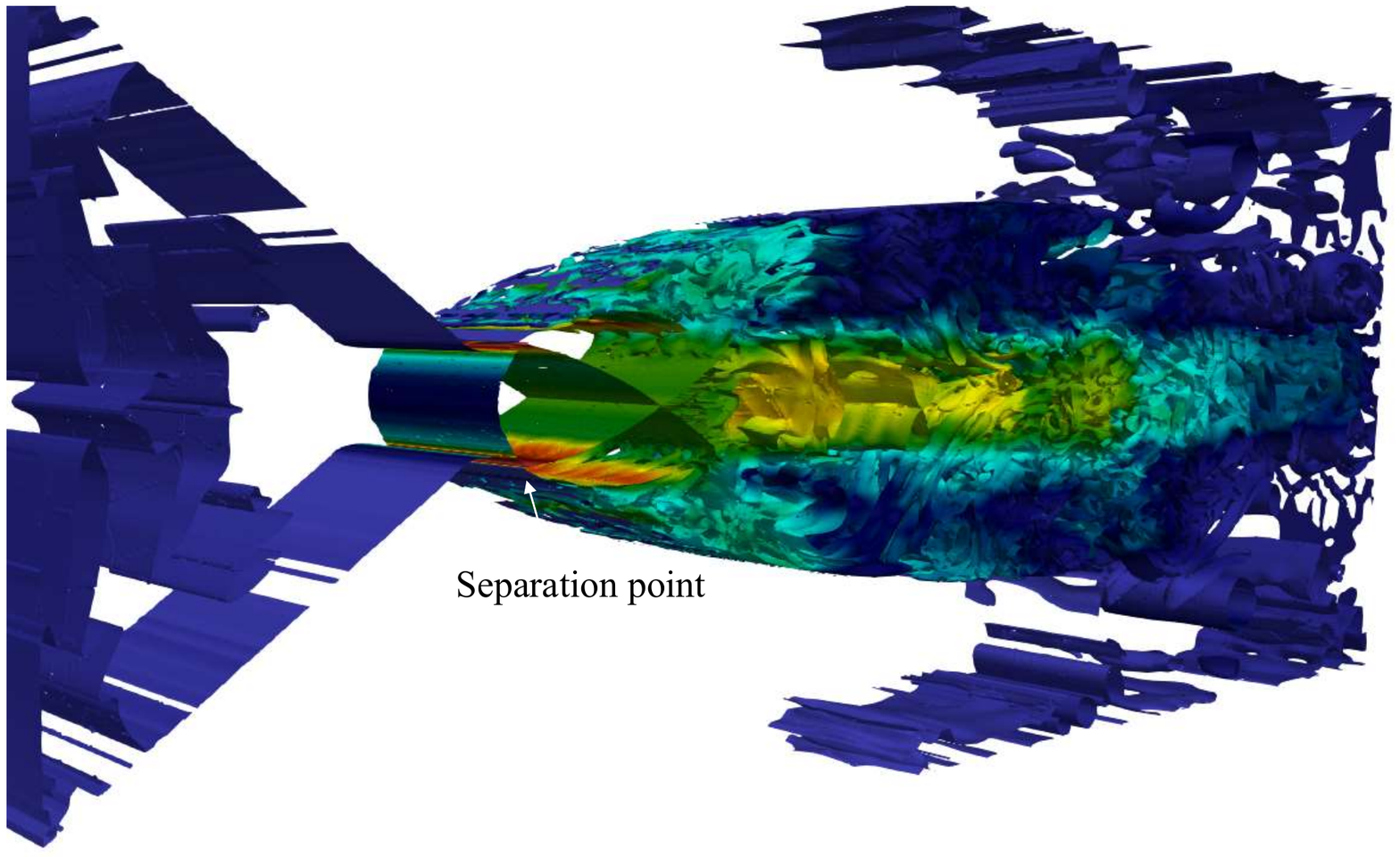}
         \caption{\label{fig:Bxprime}}
         
     \end{subfigure}
     \vspace{0.01cm}
     \begin{subfigure}[b]{\columnwidth}
         \centering
         \includegraphics[trim = 1cm 3cm 1cm 3cm, width=\textwidth]{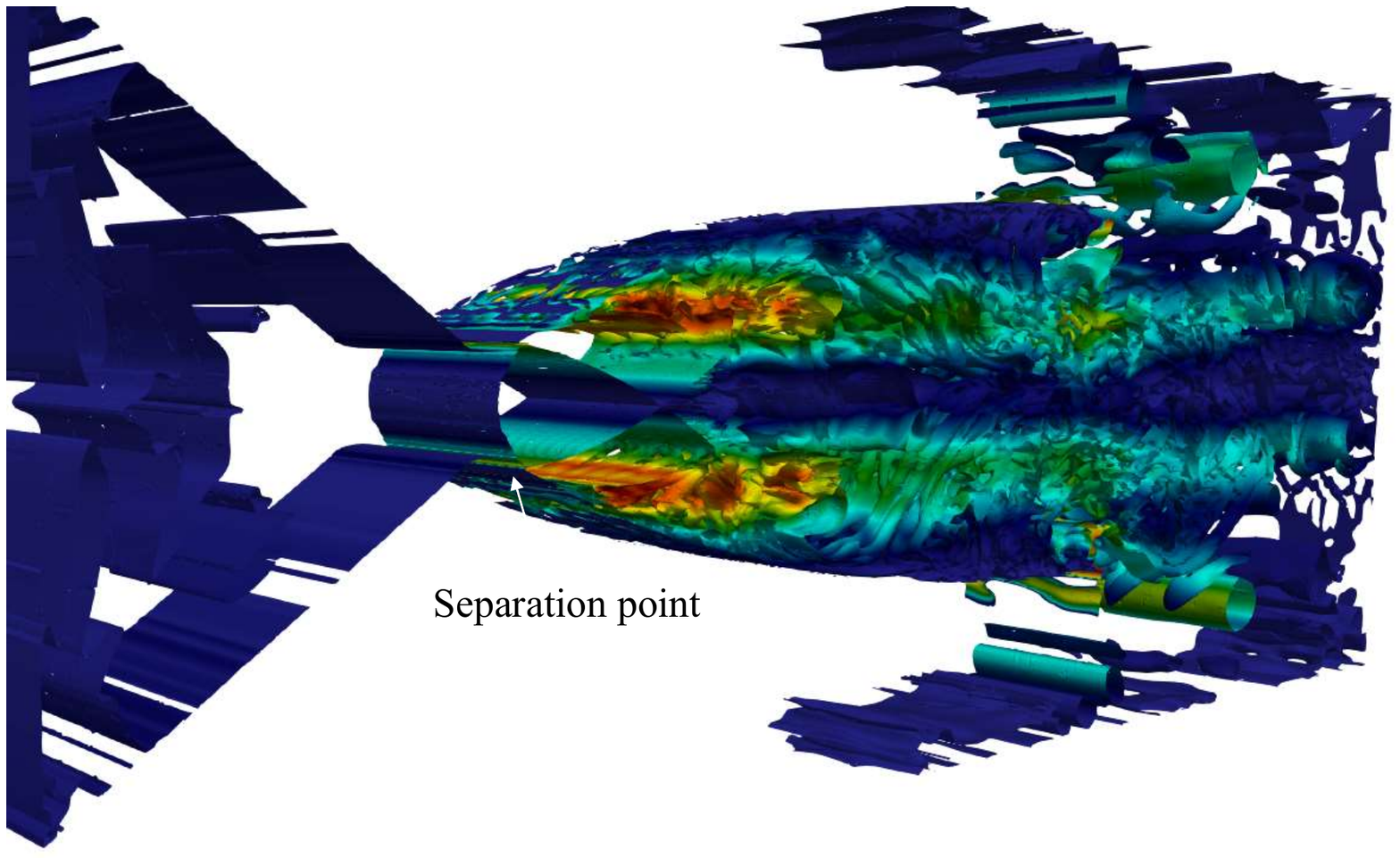}
         \caption{\label{fig:Byprime}}
         
     \end{subfigure}
        \caption{ Iso-volume of the three dimensional Q-criterion, (a) coloured by the streamwise Reynolds stress $\tilde{u^{\prime 2}}$. (b) coloured by the wall-normal Reynolds stress, $\tilde{v^{\prime 2}}$.  }
        \label{fig:reynoldsstress}
\end{figure*}

Before analysing the unsteadiness frequencies in the flow, we first study the turbulence transport mechanisms more closely. To do this, we also plot the Q-criterion coloured by the streamwise and wall-normal variance of velocity (Fig.~\ref{fig:reynoldsstress}), $\tilde{u'}^2$  and $\tilde{v'}^2$. It can be clearly observed that the amplification of the streamwise component of the Reynolds stress always precedes the wall-normal one, which develops at the later stages within the shear layer. Such observations of the delayed evolution between the two components have already been made in numerous canonical and external flow SBLI configurations 
\citep{dupont2005space,dupont2006space,pirozzoli2006direct,priebe2012low,fang2020turbulence}, and emphasises the similarities between the SBLI in nozzles and other configurations despite significant geometrical differences in the flow domain. Thus, we here provide a brief qualitative explanation for this through the use of the turbulence production term in the Reynolds transport equations \citep{cebeci2012analysis}:
\begin{equation}
\begin{split}
P_K=-\bar{\rho}\left\langle u_i^{\prime \prime} u_j^{\prime \prime}\right\rangle \frac{\partial\left\langle u_i\right\rangle}{\partial x_j}=\underbrace{-\bar{\rho}\left\langle u^{\prime \prime} v^{\prime \prime}\right\rangle\left(\frac{\partial\langle u\rangle}{\partial y}+\frac{\partial\langle v\rangle}{\partial x}\right)}_{P_s}
\\ 
\underbrace{-\bar{\rho}\left\langle u^{\prime \prime} u^{\prime \prime}\right\rangle \frac{\partial\langle u\rangle}{\partial x}}_{P_x} \underbrace{-\bar{\rho}\left\langle v^{\prime \prime} v^{\prime \prime}\right\rangle \frac{\partial\langle v\rangle}{\partial y}}_{P_y}
\\
- \bar{\rho} \langle u''  w''\rangle \left(\frac{\partial \langle u \rangle}{\partial z}  + \frac{\partial \langle w \rangle}{\partial x}\right)- \\ \bar{\rho} \langle v''  w'' \rangle \left(\frac{\partial \langle v \rangle}{\partial z} 
+ \frac{\partial \langle w \rangle}{\partial y}\right)- \bar{\rho} \langle w''  w'' \rangle \frac{\partial \langle w \rangle}{\partial z}
\end{split}
\end{equation}
where we have included the spanwise contributions of the production term for completeness. They are however, often negligible especially at separation considering the turbulence amplification is primarily driven by the streamwise and wall-normal fluctuations. At separation, the streamwise pressure gradient is highly adverse, resulting in always that the strain rate in the streamwise direction, $d\langle u \rangle /dx$ amplifying first \citep{smits2006turbulent}. However, the term contributing to the wall-normal variance $P_y$, is a negative value, since  $d\langle u \rangle /dy$ is less than zero in adverse pressure gradient flows. This altogether implies that turbulence production in the streamwise direction will always be amplified first, and at later stages viscous shear stresses dominate the mixing layer flow and the wall-normal gradients will again evolve once the flow becomes fully-developed. 

\subsection{Frequency spectra }

\begin{figure*}
\includegraphics[width = \columnwidth]{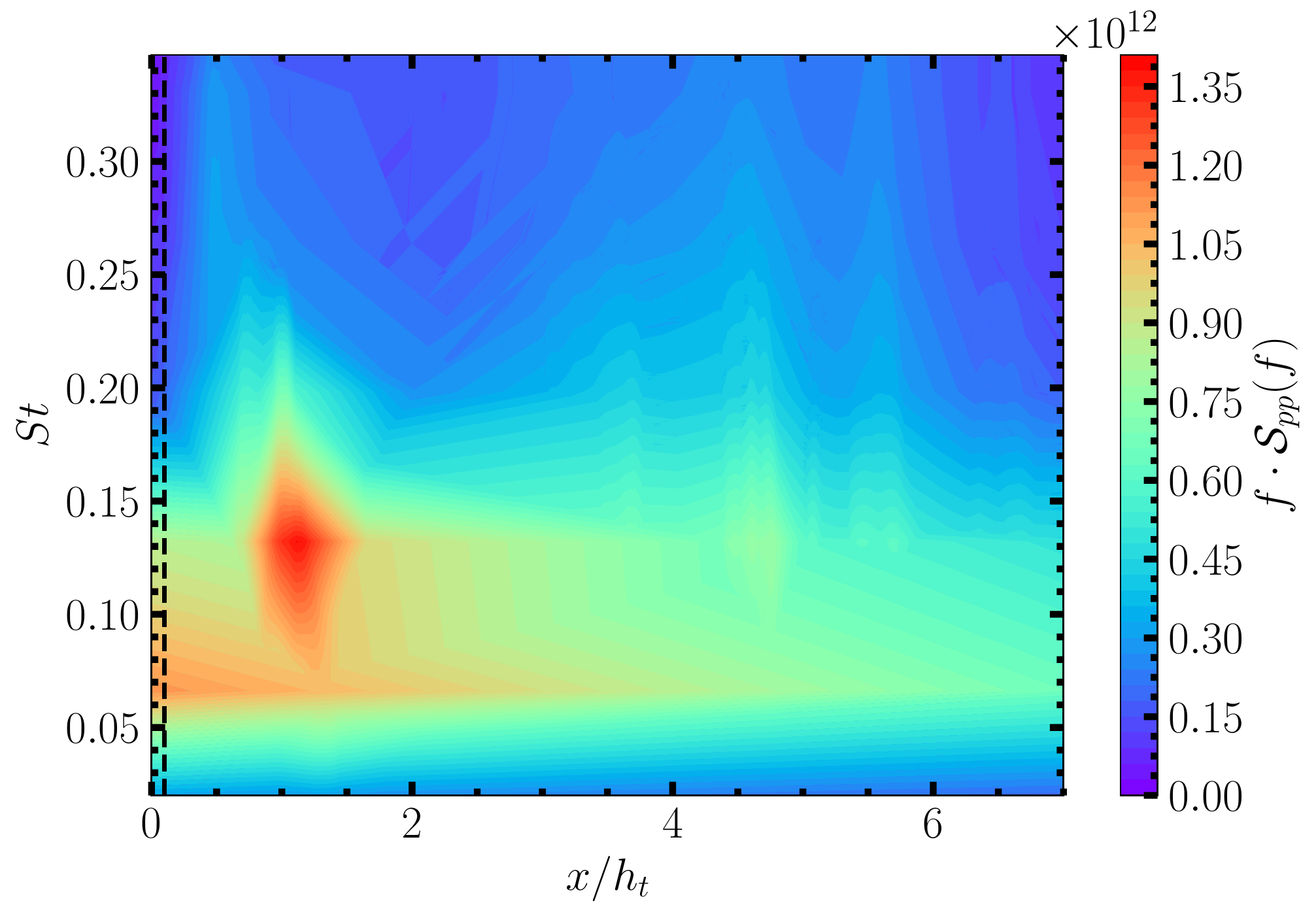}
\caption{\label{fig:spectra} Pre-multiplied power spectra of the wall pressure perturbations $f \cdot \mathcal{S}_{pp}(f)$ as functions of the non-dimensional streamwise coordinate, $x/h_t$ and the Strouhal number $St = fh_j/U_j$. Dashed black line indicates the location of incipient flow separation from the nozzle wall. }
\end{figure*}
Here we examine the frequency spectra and identify the wall-pressure unsteadiness in the nozzle. Fig.~\ref{fig:spectra} displays the nozzle wall-pressure frequency spectra as a function of streamwise coordinate. The power spectral density is estimated via Welch's periodogram method, where the synchronous time-varying data across the nozzle wall are split into 12 FFT (Fast-fourier transform) intervals for adequate frequency resolution. A Hanning window is further used to reduce spectral leakage across different intervals. The wall-pressure signals clearly indicate high amplitude signatures at two distinct Strouhal numbers $St \approx 0.06$ and $St \approx 0.15$, which are seemingly the low and high unsteadiness frequencies of SBLI, where the broad low frequency footprint begins exactly at the separation point near the nozzle throat located at $x/h_t = 0$, and the fluctuating energy remains further downstream up until about $x/h_t \approx 4.5$. The higher frequency unsteadiness of $St \approx 0.08 - 0.22$, occurs slightly downstream and is related to the turbulent shear layer forming at the jet boundary discontinuity, which is dominated by developing coherent structures advected through the mean flow. This value of $St$ is surprisingly very similar to that observed in prior TOC (Vulcain) and TIC nozzle studies (e.g., \citet{dumnov1996unsteady,jaunet2017wall,bakulu2021jet}), where an $St = 0.2$ energy peak was also observed with fully-expanded Mach number $M_j = 2.09$, based on the analogous fully-expanded jet diameter $D_j$ from the isentropic relation of \citet{tam1988shock}.

The overall qualitative features of the wall-pressure spectra are also very similar to those obtained in prior works \citep{jaunet2017wall,martelli2017detached,martelli2020flow,zebiri2020shock,bakulu2021jet}, where the salient features of the wall-pressure PSD shares many qualitative similarities with the canonical OSWBLI configurations (e.g.,~\citet{dupont2006space}), even though the flow boundaries here are fundamentally very different in nature with measurements from compression ramps, blunt fins etc. 

\begin{figure*}
     \centering
     \begin{subfigure}[b]{\columnwidth}
         \centering
         \includegraphics[width=\textwidth]{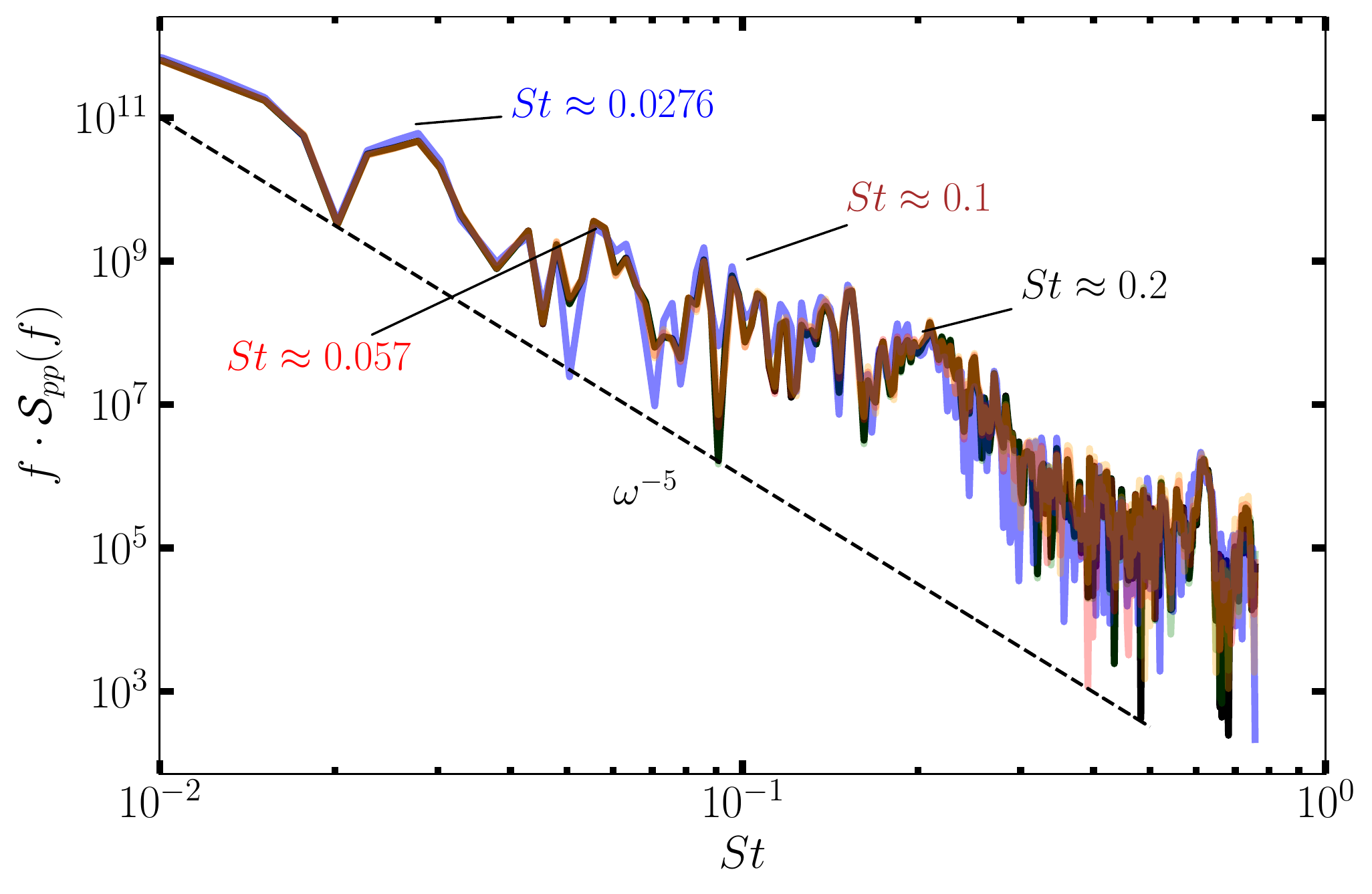}
         \caption{\label{fig:sepspec}}
     \end{subfigure}
     \begin{subfigure}[b]{\columnwidth}
         \centering
         \includegraphics[width=\textwidth]{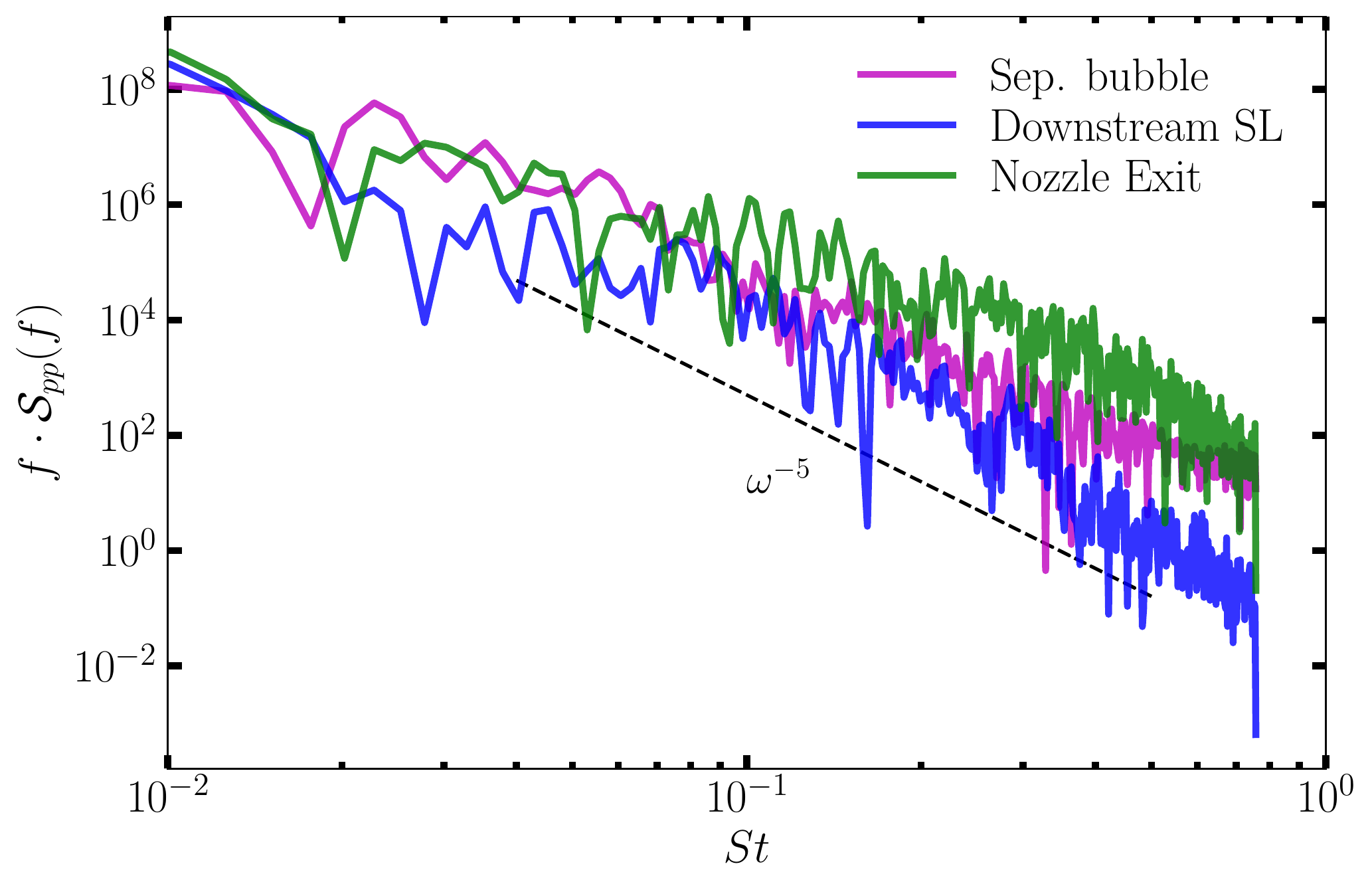}
         \caption{\label{fig:morespec}}
         
     \end{subfigure}
        \caption{(a) Frequency power spectrum of the pressure perturbations at the streamwise station $(x/h_t, y/h_t) = (0.5,0.5)$, directly located at the separation point. Several power spectra are taken based on cells corresponding to this location, in order to fully resolve all the temporal dynamics in this region. (b) Frequency spectrum at various streamwise locations in the separation bubble, downstream shear layer, and nozzle exit. All diagrams are fitted with a $\omega^{-5}$ scaling consistent with that observed in canonical SWBLI (e.g., \citet{pirozzoli2010direct,bernardini2011wall,fang2020turbulence}).  }
        \label{fig:freqspec}
\end{figure*}
There are however, some subtle differences we observe here as compared to numerous prior high Reynolds number overexpanded nozzle studies (cf. \citet{baars2012wall,martelli2020flow,bakulu2021jet}), with high jet Mach numbers. The footprint associated with $St \approx 0.15$, usually persists much further downstream since the turbulent intensity of the compressible Kelvin-Helmholtz (K-H) mode in the stratified shear layer decays exponentially, allowing the sustainment of pressure perturbations that radiate outwards and continually impinges on the nozzle wall. However, since the TBL in our simulation separates significantly earlier than in the prior works, wherein the NPR typically ranged between $5 - 50$, or had smaller nozzle aspect ratios \citep{papamoschou2009supersonic, johnson2010instability}, these features are seemingly no longer visible on the nozzle wall, and instead only exist slightly downstream of the separation point. It can also be attributed to the thickening of the shear layer, which dampens the frequencies observed with increasing streamwise separation, as reported in the experiments and simulations by \citet{jaunet2017wall}.   

Due to these discrepancies in the wall-pressure spectrum, we resort to analysing frequency spectra at specific spatial locations in order to further isolate the dominant eigenfrequencies of the flow. In particular, we focus on locating the shock unsteadiness frequency. Fig.~\ref{fig:freqspec} displays several frequency spectra of the wall-pressure perturbations at individual points. Fig.~\ref{fig:sepspec} is the spectra obtained based on points directly at the separation point, where it can be seen that more detailed peaks and signatures can be observed. The separation between the lowest and highest frequencies with high spectral signature is about a decade apart. Similar ranges have also been observed by \citet{zebiri2020analysis}. Moreover, \citet{baars2012wall} and \citet{zebiri2020shock} both suggested that the low frequency tone could correspond to \citet{zaman2002investigation}'s transonic resonance, which can be modelled as:
\begin{equation}
    St_{\textrm{res}} = \frac{a_{\infty}h_j(1 - M_{\textrm{NE}}^2)}{4LU_j} 
\end{equation}
where $a_{\infty}$ is the ambient speed of sound and $L$ is the streamwise length between the separation point and the nozzle exit, and $M_{\textrm{NE}}$ is the Mach number at the nozzle exit. For our nozzle, it corresponds to $St_{\mathrm{res}} \approx 0.024$, a value very close to that observed in the spectra. We further note the existence of other fluctuating signatures, which align well with that obtained in the wall-pressure spectra earlier. Also, frequency pressure spectra within the separation bubble demonstrates a rather rapid cascade, with only a peak scale at the low frequencies, with Strouhal numbers of order $St \sim \mathcal{O}(10^{-2})$. These are related to the separation bubble dynamics. As observed in \citet{olson2011large,olson2013mechanism} and \citet{zebiri2020shock}, the separation bubble is dominated with low frequencies that are driven by the pressure imbalance generated from the separation point and the nozzle lip. This is thus correlated with the existence of downstream and upstream propagating waves due to the dilatation and contraction of the separation bubble \citep{piponniau2009simple}.

The other power spectra computed for the downstream shear layer and nozzle exit, contain signatures that persist up to $St \sim \mathcal{O}(10^{-1})$, thus confirming that the intermediate to high frequency range fluctuating energies are related to vortex shedding in the detached layer, as well as mixing layer unsteadiness, respectively; as similarly observed by \citet{zebiri2020analysis} and \citet{zebiri2020shock}. Furthermore, all spectrs seemingly also confirm the $\omega^{-5}$ scaling observed in canonical SWBLIs\citep{pirozzoli2010direct,bernardini2011wall,fang2020turbulence}, despite the differing regions sampled. This strongly suggests the possibility of a universal scaling for pressure spectra in SBLIs, up to small-scale intermittency corrections.
\begin{figure}
\includegraphics[width = \columnwidth]{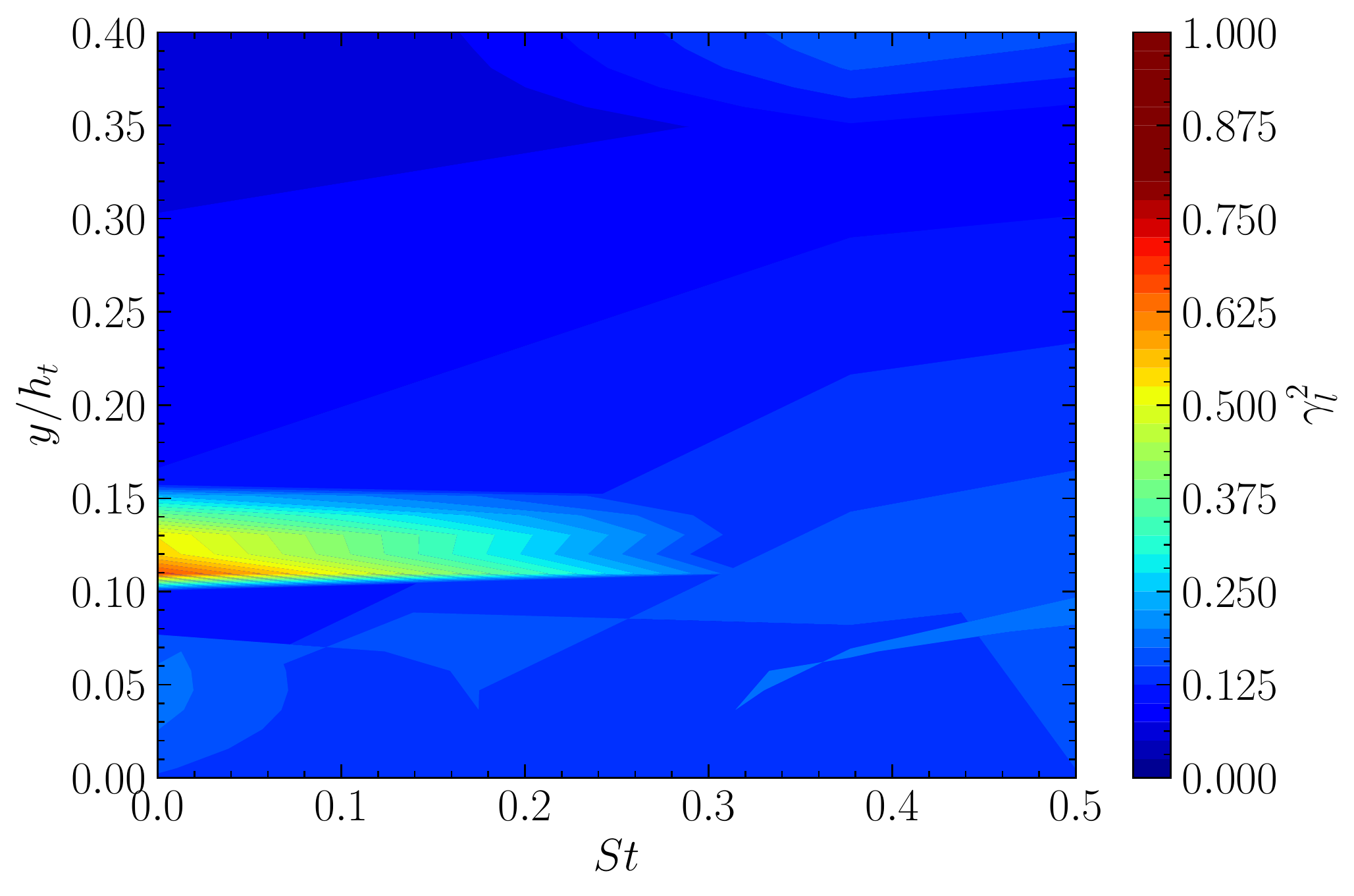}
\caption{\label{fig:coherence} Magnitude squared linear coherence spectrum between pressure perturbations at two streamwise stations, $x/h_t = 2 $, and $x/ h _t =5$, as a function of the wall-normal coordinate. Clearly, only the low Strouhal number tones remain correlated.    }
\end{figure}

In order to further validate that the spectral signatures in each region are fundamentally uncorrelated with each other, where the intermediate to high frequencies are intrinsically linked to the development of coherent structures at later time evolutions, we also plot the a magnitude squared linear coherence spectrum of the pressure perturbations (normalised cross spectral density), between two spatial locations, as a function of the wall normal-coordinate $y$ (Fig.~\ref{fig:coherence}). It can be seen that only the low frequency dynamics are highly correlated, and also concentrated only within the separation bubble. This further suggests that the low frequency tones play a fundamental role in the dynamics and tonal flow behaviour of the flow system, and also strongly indicates the likelihood of downstream mechanisms being involved in the sustainment of low frequency unsteadiness.

Furthermore, the observation that the high frequencies are fundamentally decorrelated with each other across different spatial locations is consistent with the prior observation that they originate from convective instabilities and vortical structures within the shear layer, and hence are naturally highly anisotropic in nature. 

\subsection{Aeroacoustic resonance}
\subsubsection{Thermal crackle noise}
\begin{figure}
\includegraphics[width = \columnwidth]{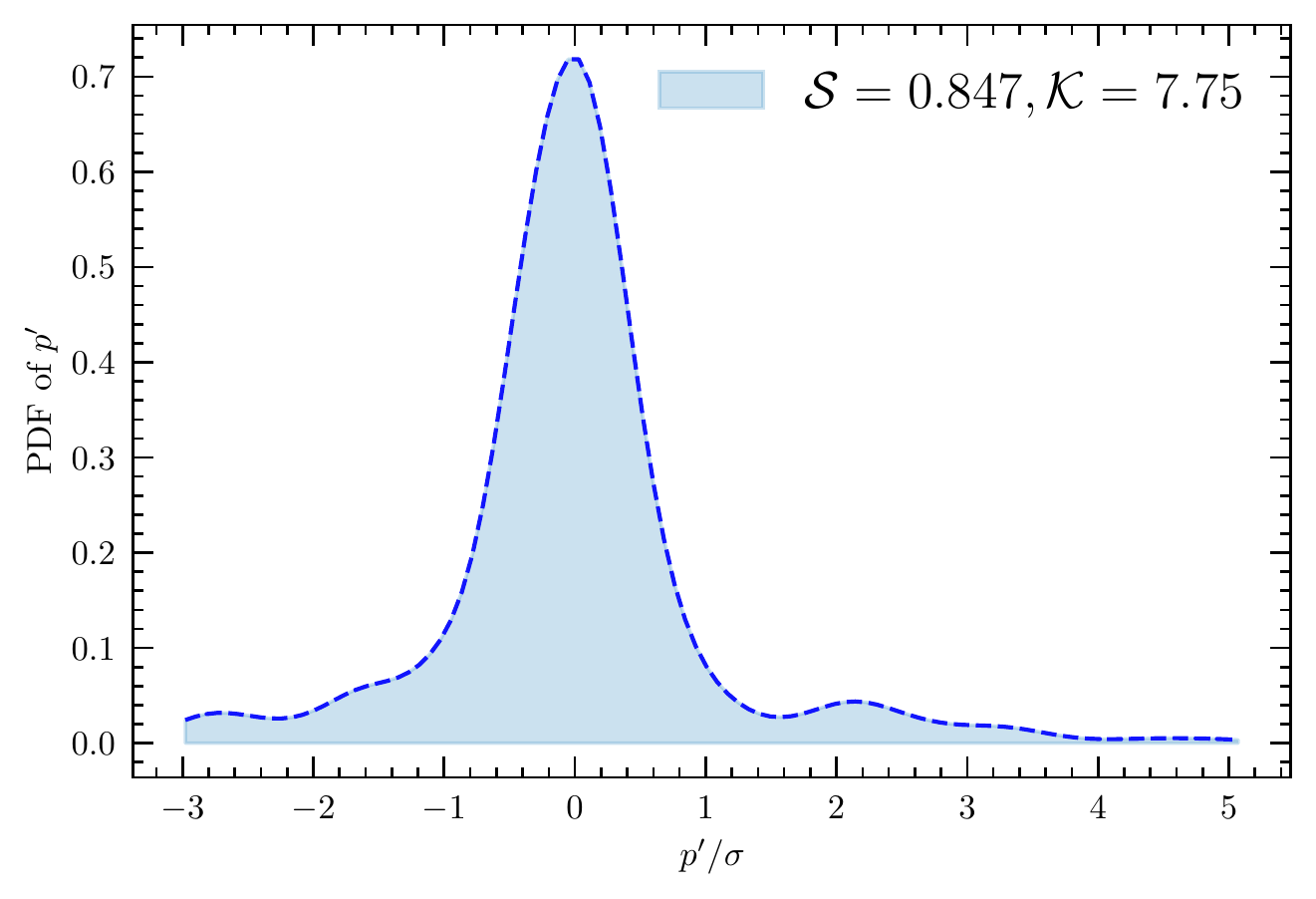}
\caption{\label{fig:pdf} Probability distribution function (PDF) of the pressure fluctuations sampled across the entire jet shear layer (at $y = 0.8 h_t$), averaged in time. Skewness ($\mathcal{S}$) and kurtosis ($\mathcal{K}$) factors indicate non-Gaussian stretched tails and spatial intermittency associated with supersonic TBL separation \citep{kistler1963fluctuating}.}
\end{figure}

Now we focus on the aeroacoustics involved with the pressure perturbations. Fig.~\ref{fig:pdf} displays the probability distribution function of the streamwise pressure fluctuations within the detached shear layer at $y = 0.8h_t$. High order moments of the PDF; namely the skewness and kurtosis factors:
\begin{equation}
\begin{aligned}
\mathcal{S}_q &=\frac{\left\langle(q-\langle q\rangle)^3\right\rangle}{\sigma^3} \\
\mathcal{K}_q &=\frac{\left\langle(q-\langle q\rangle)^4\right\rangle}{\sigma^4}
\end{aligned}
\end{equation}
indicate non-Gaussian behaviour with stretched tails, suggesting overall spatial intermittency within the jet shear layer, where $\mathcal{S} \approx 0.847^{+ 0.002}_{-0.001}$ and $\mathcal{K} \approx 7.75 \pm 0.06$. We note here that while it is common to compute higher order moments to identify the so-called thermal crackle noise \citep{williams1975crackle,nichols2013crackle,mora2014impact} associated with heated unseparated jets, where $\mathcal{S} > 0.3$ is a representative cutoff where thermal crackle noises are likely to occur \citep{nichols2013crackle,nichols2013source,mora2014impact,chen2021flow}; however, such measures do not likely hold for an over-expanded separated jet, since we observe high intermittency even with standard cold walls. These spatial intermittencies as found in the higher order moments are thus likely more related to the convective instabilities within the detached shear layer \citep{kistler1963fluctuating,na1998structure,bernardini2011wall}, rather than specific kinds of noise generation. 
\subsubsection{Estimation of the convection velocity}
In order to fully educe the dynamics of the pressure fluctuations in a scale-dependent manner, as well as to quantify the velocities of either upstream or downstream propagating waves,  we propose to utilise the frequency-dependent convective velocity model as proposed by \citet{renard2015scale}. Much in contrast to the space-time correlation function method \citep{favre1983turbulence,na1998structure,bernardini2011wall}, which utilises the maximisation of the correlation coefficient for a given time and space lag, this formulation allows for the decomposition of each velocity component in terms of its characteristic frequency:
\begin{equation}
    u_c(f) = \frac{-2 \pi fG_{pp}}{\textrm{Im}(G_{pp^{\prime}})}
\end{equation}
where $G_{pp}$ is the auto-spectrum between the wall-pressure fluctuations and $G_{pp^{\prime}}$ is the cross spectrum between pressure and its streamwise derivative $dp/dx$. This method was also used by \citet{martelli2020flow} in order to obtain a frequency dependent convective velocity. 

\begin{figure*}
     \centering
     \begin{subfigure}[b]{\columnwidth}
         \centering
         \includegraphics[width=\textwidth]{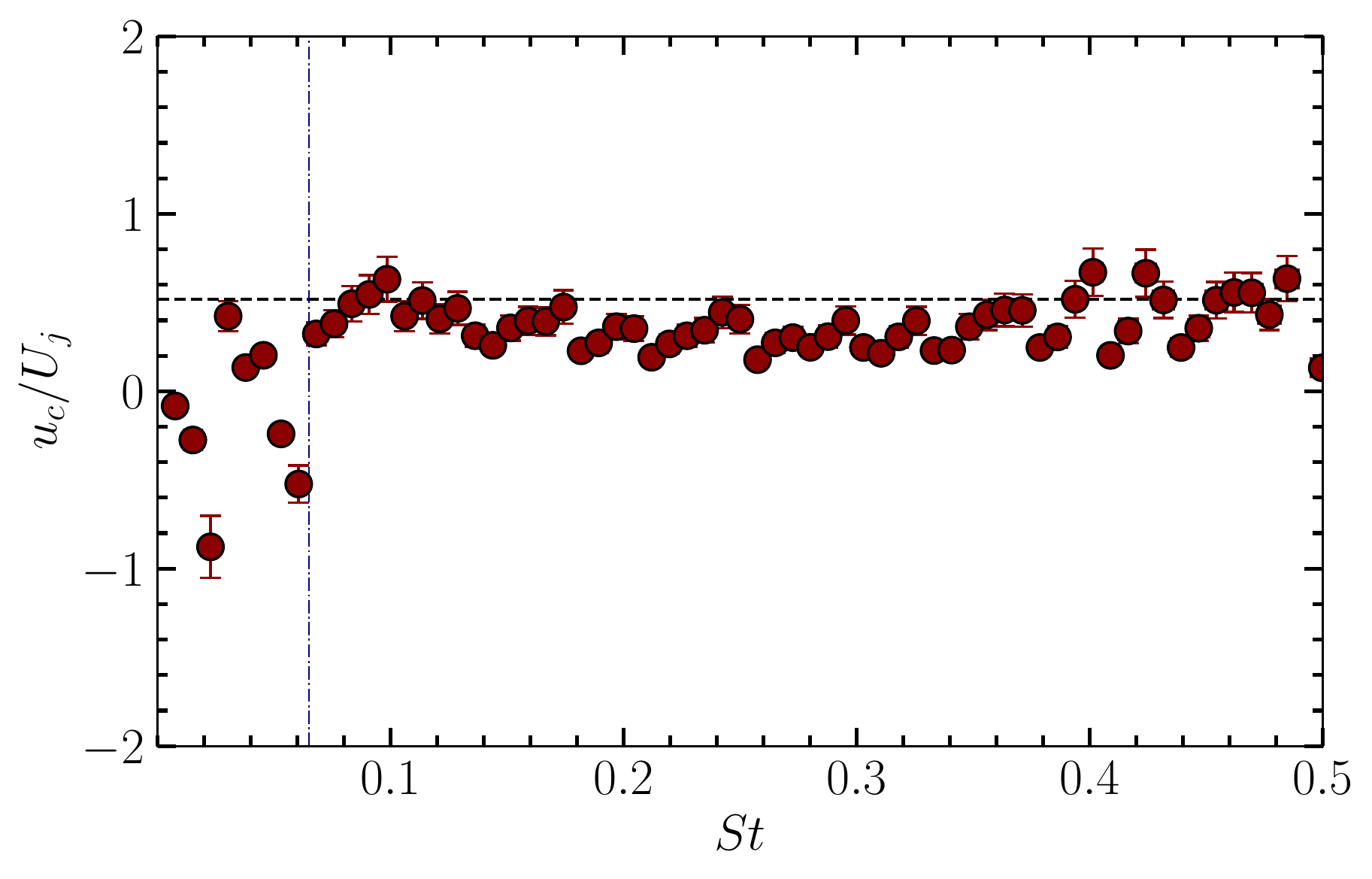}
         \caption{\label{fig:ucuj}}
     \end{subfigure}
     \begin{subfigure}[b]{\columnwidth}
         \centering
         \includegraphics[width=\textwidth]{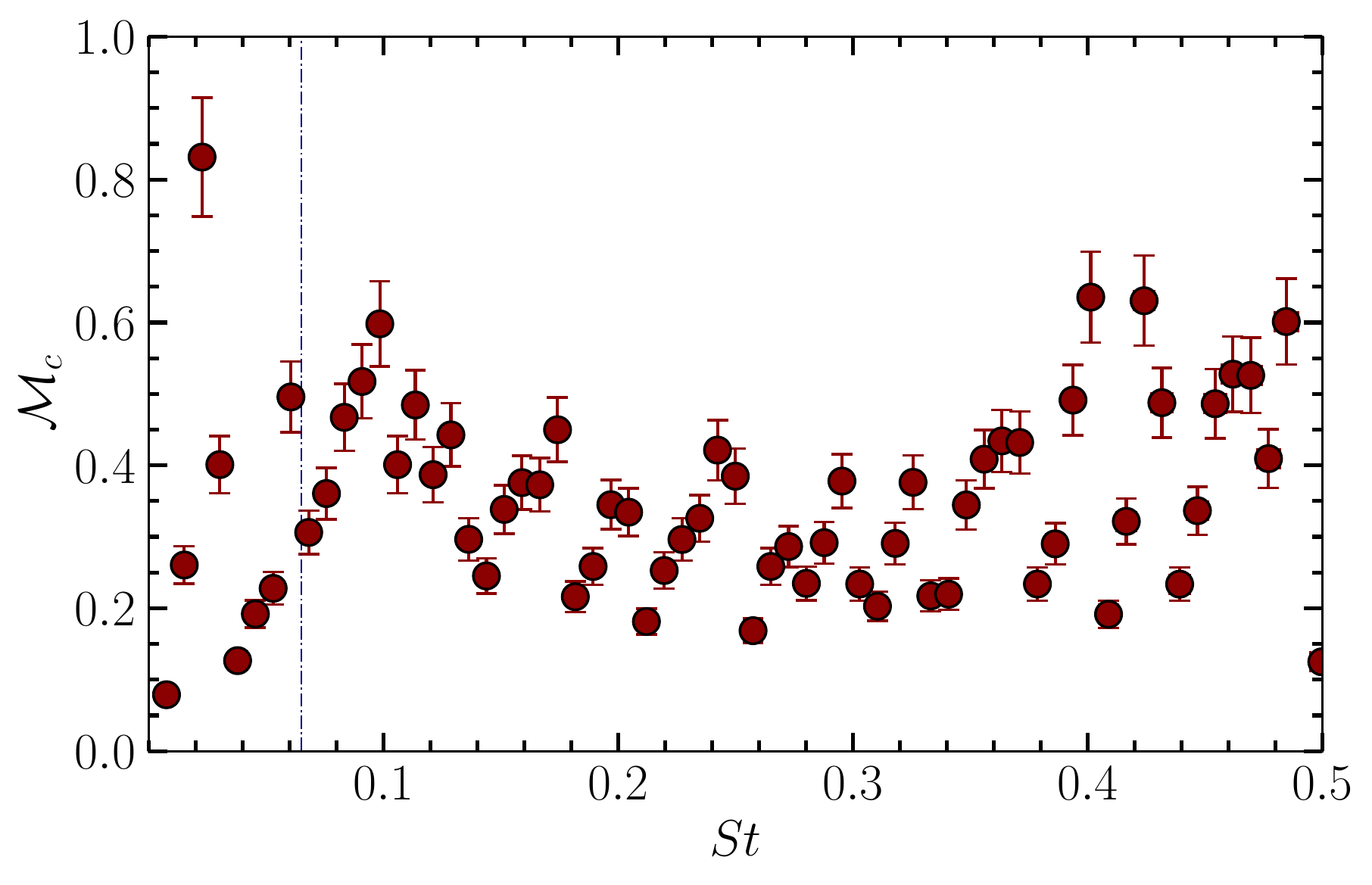}
         \caption{\label{fig:convecMc}}
         \end{subfigure}
        \caption{(a) Convective velocity normalised by fully expanded jet velocity $U_j$ as a function of Strouhal number, horizontal dashed line indicates the expected convective velocity $u_c \approx 0.52 U_j$ assuming isentropic, parallel matched streams \citep{papamoschou1988compressible}. Vertical dashed line indicates region in which upstream propagating waves exist. (b) Convective Mach number, computed as $M_c = u_c/\sqrt{\gamma R T_j}$, the same vertical line is drawn. Values are computed at location x = $4h_t$ on the wall. }
        \label{fig:uc}
\end{figure*}

Also, as pointed out by \citet{renard2015scale}, the accuracy of the model can be sensitive to how the local streamwise derivative of the fluctuation is evaluated, which can be enhanced if high-order finite difference approximations with more cell stencils are used. Therefore, here we use the five-point stencil method based on the fourth-order central difference scheme: 
\begin{equation}
\frac{dp}{dx}\rvert_{x_i} \approx \frac{-p_{i + 2 } + 8p_{i +1 } - 8p_{i-1}  + p_{i -2}}{12 \Delta x} 
\end{equation}
where the derivative at the grid point $x_i$ with index $i$, depends on the value obtained at four neighbouring grid points. This yields a discretisation error of $\mathcal{O}(\Delta x)^4$ by the Richardson extrapolation.
We also compare this result to the case of two isentropic, parallel-matched streams \citep{papamoschou1988compressible,gojon2016investigation}:
\begin{equation}
u_c = \frac{U_j}{a_j/a_{\infty} +1 } \approx 0.52 U_j
\end{equation}
Fig.~\ref{fig:uc} displays the convective velocity and convective Mach numbers computed, and confirms the possibility of upstream propagating disturbances with negative phase velocities at the low frequency range. This also aligns with the observation of \citet{martelli2020flow}, who applied this method and found that upstream propagating waves are only supported for low to intermediate $St$ numbers. The convective Mach number plot (Fig.\ref{fig:convecMc}), also crucially shows that any upstream propagating waves must be subsonic. Thus, they can only travel through the separation bubble, and never within the core flow, supporting the idea that the separation bubble plays a crucial role in maintaining the aeroacoustic feedback loop mechanism.

\subsubsection{Screech mechanism}

\begin{figure*}
     \centering
     \begin{subfigure}[b]{\columnwidth}
         \centering
         \includegraphics[width=\textwidth]{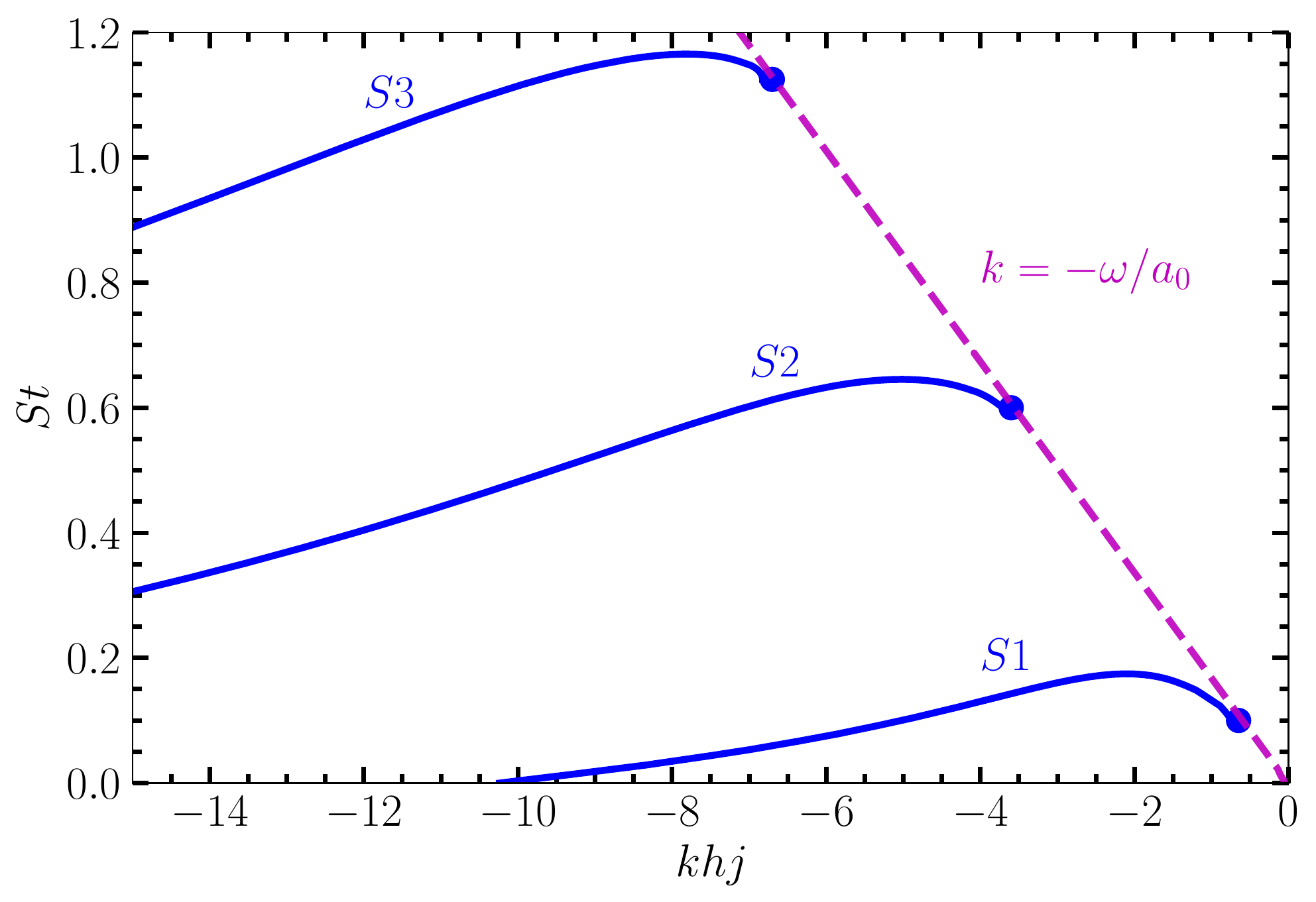}
         \caption{\label{fig:ucuj}}
     \end{subfigure}
     \begin{subfigure}[b]{\columnwidth}
         \centering
         \includegraphics[width=\textwidth]{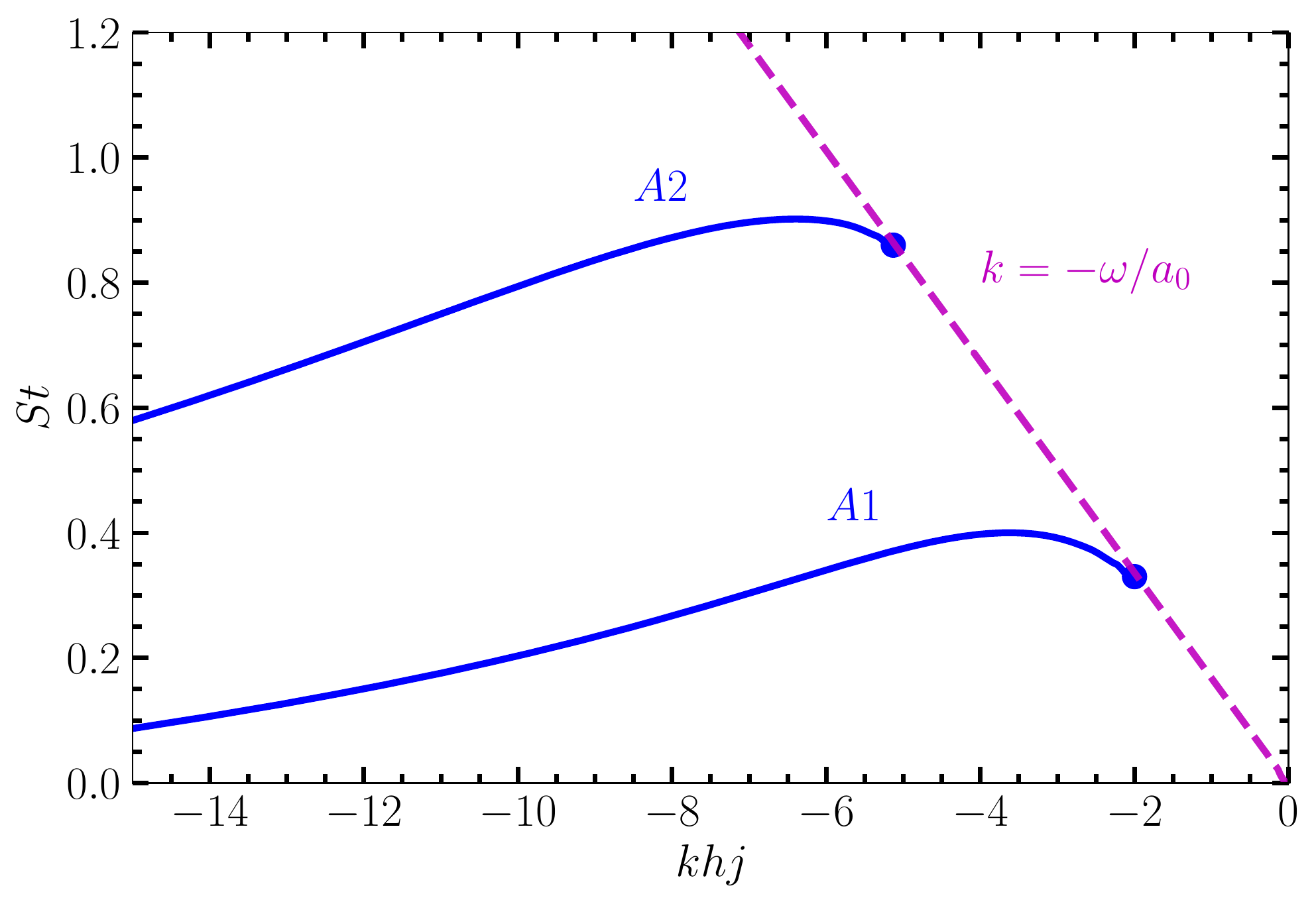}
         \caption{\label{fig:asymmodes}}        
     \end{subfigure}

 \caption{\label{fig:modes}Dispersion relations for the supported upstream propagating guided jet modes at $M_j = 1.046$ via the vortex sheet theory (a) the first three symmetric modes, (b) the first two anti-symmetric modes, a cutoff frequency exists at $k = -\omega/a_0$ as indicated.  }
\end{figure*}
\begin{figure*}
\centering
\includegraphics[width = \columnwidth]{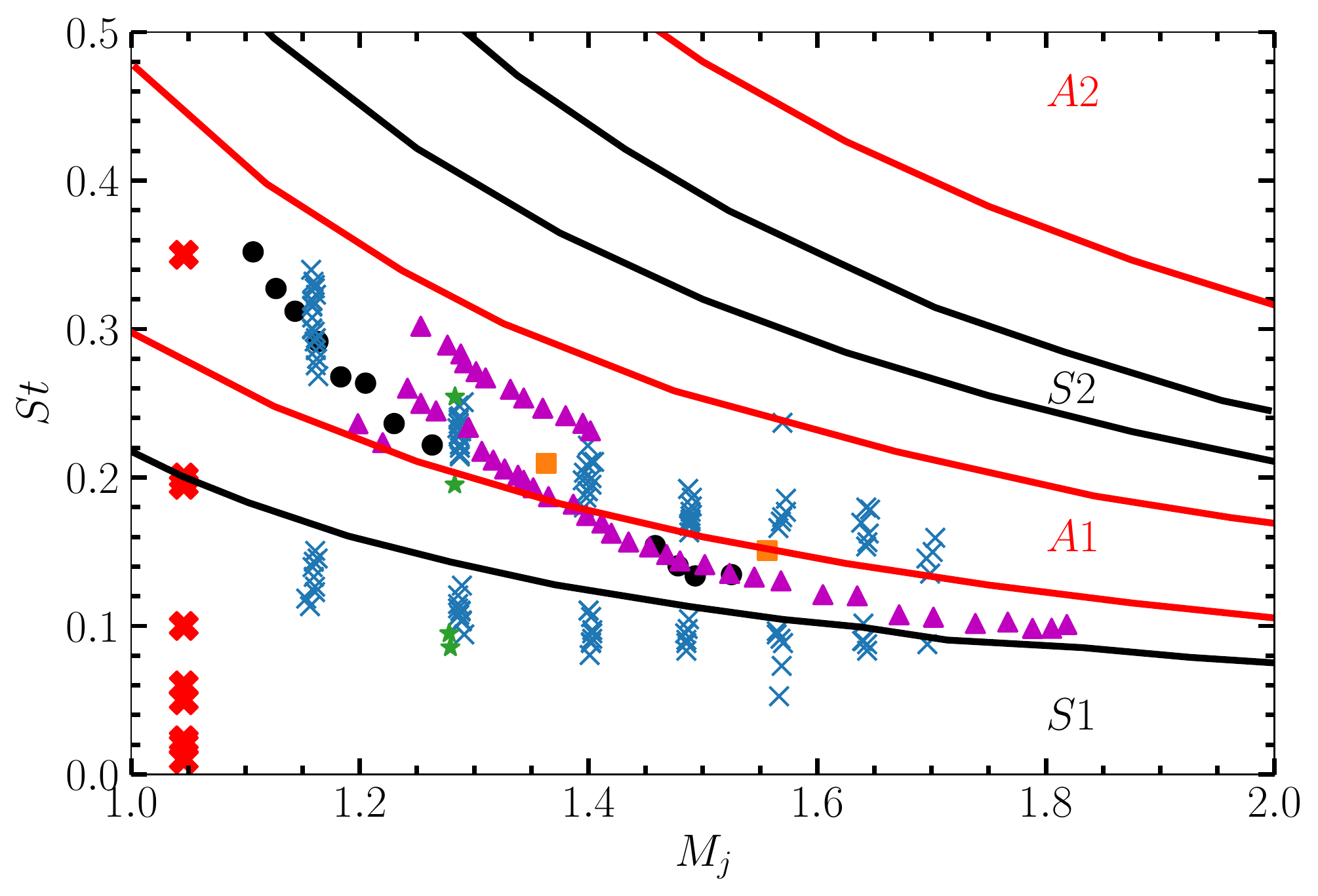}
\caption{\label{fig:jetcompare} Plot of the allowable frequency ranges for the upstream-propagating guided jet mode via the vortex sheet analysis, as a function of the fully-expanded jet Mach number. The annotated regions $S1$, $A1$, $S2$, and $A2$ indicate areas where symmetric and anti-symmetric modes can exist. With discrete tones observed in the present DDES, red $\times$, LES from \citet{gojon2016investigation} $\star$ , LES from \citet{gojon2019antisymmetric} $\square$, experiments from  \citet{panda1997underexpanded} $\bigcirc$, \citet{raman1997screech} $\triangle,$ and \citet{tam1992impingement} blue $\times$. }
\end{figure*}
The intermediate range of frequencies observed within the pressure spectra are typically related to the mixing layer unsteadiness, \citet{jaunet2017wall} and \citet{martelli2020flow} suggested that it can also induce a screech like mechanism. They applied the \citet{tam1986proposed} model and found good agreement with the dominant tones recorded in their axisymmetric jets. Here we use the rectangular jet variant of the model, where by the weakest-link theory of \citet{tam1986proposed}, \citet{tam1988shock} derived an approximate model for the screech mechanism suitable for non-axisymmetric jets. The model suggests that screech production is induced by an aeroacoustic feedback loop where downstream propagating waves in the turbulent shear layer interacts with subsequent quasi-periodic shock cells, which in turn triggers upstream travelling acoustic radiation that re-enter back into the shear layer. The approximate formula provided is: 
\begin{equation}
\begin{aligned}
St_{screech} = \frac{f h_j}{u_j}=0.7\left(\frac{1+\frac{\gamma-1}{2}}{1+\frac{\gamma-1}{2} M_j^2}\right)^{(\gamma+1) / 2(\gamma-1)} 
\\ \times
\frac{M_j}{\left[2\left(M_j^2-1\right)\left[1+\frac{0 \cdot 7 M_j}{\left(1+\frac{\gamma-1}{2} M_j^2\right)^{1 / 2}}\right]\right]}
\end{aligned}
\end{equation}
which for the current nozzle yields $St_{screech}  \approx 0.353 $. Such a frequency was not detected in the spectral analysis, which may suggest that the formula may be possibly inaccurate for very low fully-expanded Mach numbers, since the underlying assumption is that the convective velocity $u_c \approx 0.7u_j$,  based on experiments by \citet{harper1974noise}. However, as found from the \citet{renard2015scale} model for convective velocity, we expect $u_c \approx 0.5u_j$, which explains the possible deviation. We note also that in the later discussion of DMD in Sec.~\ref{sec:dmd}, we found a strong frequency corresponding to $St \approx 0.35$. However, this may not be related exactly to the screech mechanism supported by acoustic waves, and might be related to other kinds of instabilities in OSWBLI within separated nozzle flowfields \citep{nguyen2003unsteadiness,verma2014effect,zebiri2020analysis,zebiri2020shock}. 

In light of the discrepancy with the weakest link theory, which assumes the upstream propagating component is a freestream acoustic wave. Here we apply the vortex sheet analysis of \citet{tam1992impingement} for the rectangular jet, which assumes that the upstream component is a neutral acoustic mode instead. The theory has been verified numerically (e.g., \citet{gojon2016investigation}), as well as experimentally (e.g., 
 \citet{edgington2018upstream}). The equations for the pairs of supported anti-symmetric and symmetric modes are:
The equations derived by \cite{tam1992impingement} for the vortex sheet analysis are:
\begin{equation}
\begin{aligned}
\frac{\left[\left(\omega-u_j k\right)^2 / a_j^2-k^2\right]^{1 / 2} \rho_0 \omega^2}{\left(k^2-\omega^2 / a_0^2\right)^{1 / 2} \rho_j\left(\omega-u_j k\right)^2} \\ -1/\tan \left\{\left[\frac{\left(\omega-u_j k\right)^2}{a_j^2}-k^2\right]^{1 / 2} h / 2\right\}=0
\end{aligned}
\end{equation}
for symmetric modes, and
\begin{equation}
\begin{aligned}
\frac{\left[\left(\omega-u_j k\right)^2 / a_j^2-k^2\right]^{1 / 2} \rho_0 \omega^2}{\left(k^2-\omega^2 / a_0^2\right)^{1 / 2} \rho_j\left(\omega-u_j k\right)^2} \\ +  \tan \left\{\left[\frac{\left(\omega-u_j k\right)^2}{a_j^2}-k^2\right]^{1 / 2} h / 2\right\}=0
\end{aligned}
\end{equation}
for the anti-symmetric modes, where $\omega$ is the angular frequency and $k$ is the wavenumber. Limiting frequencies exist at $k = -\omega/a_0$ as shown earlier, these correspond to the lower limit of the possible Strouhal numbers. The upper limits are found using the maximum points in the dispersion relation plots for each mode. 

Fig.~\ref{fig:modes} displays the allowable asymmetric and symmetric modes for the current jet, along with the cutoff frequencies where no modes are supported for the upstream propagating component. Based on the discrete tones we observed from the \citet{renard2015scale} model, as well as the intermediate frequencies in our spectra that is typically associated with screech, we perform an extensive comparison to data obtained in numerous other studies, as shown in Fig.~\ref{fig:jetcompare}, where we plot the allowable frequencies of the upstream propagating modes as a function of jet Mach number. It can be clearly seen that all the upstream propagating tones we observed, fit within the vortex sheet model. This suggests, for the first time, that even the over-expanded separated jet may support guided jet modes for closure of the screech loop.

\begin{figure}
     \centering
     \begin{subfigure}[b]{\columnwidth}
         \centering
         \includegraphics[trim = 0cm 0cm 0cm 1cm,width=\textwidth]{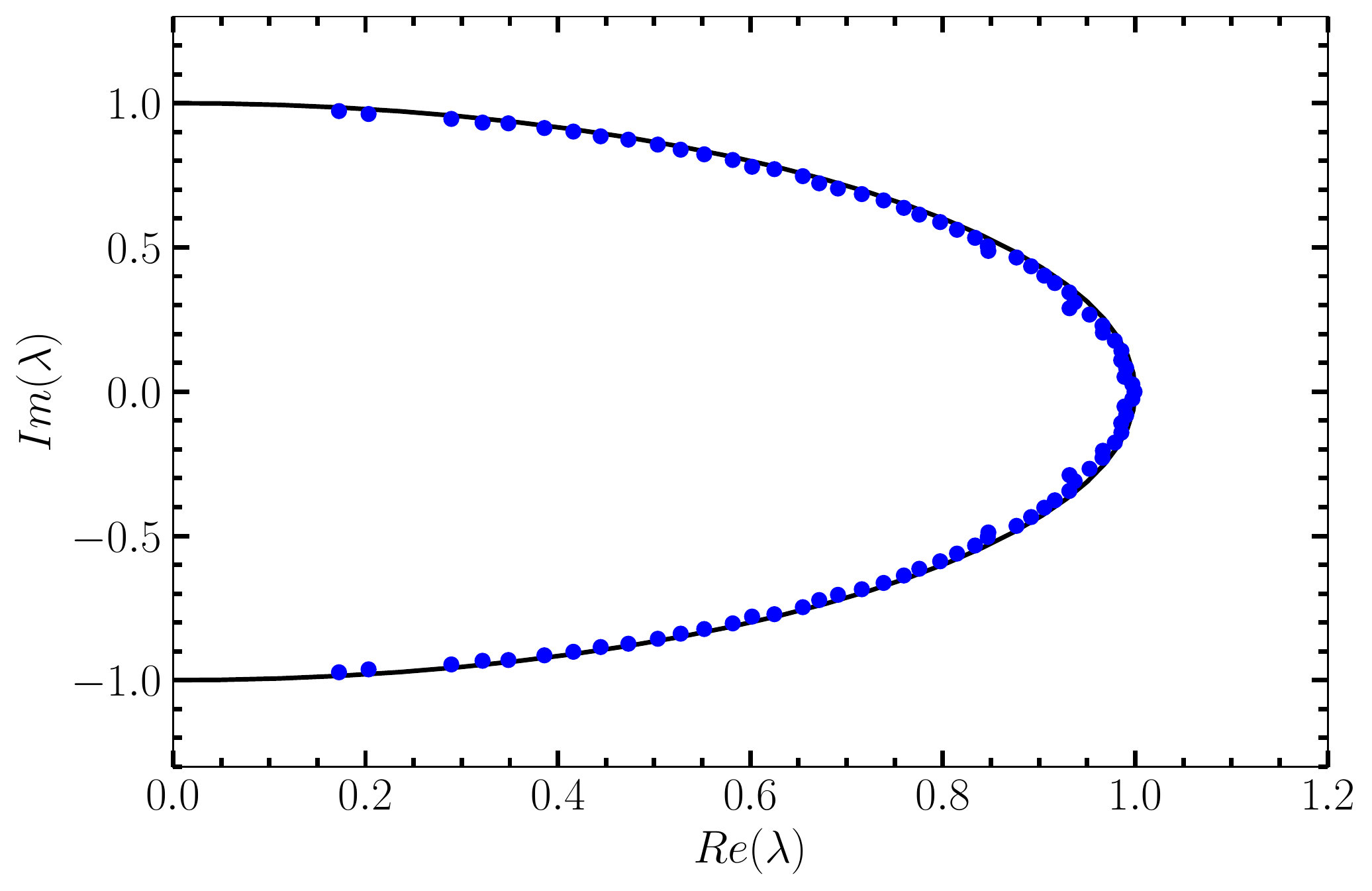}
         \caption{\label{fig:dmdeigenvalue}}
     \end{subfigure}
     \begin{subfigure}[b]{\columnwidth}
         \centering
         \includegraphics[width=\textwidth]{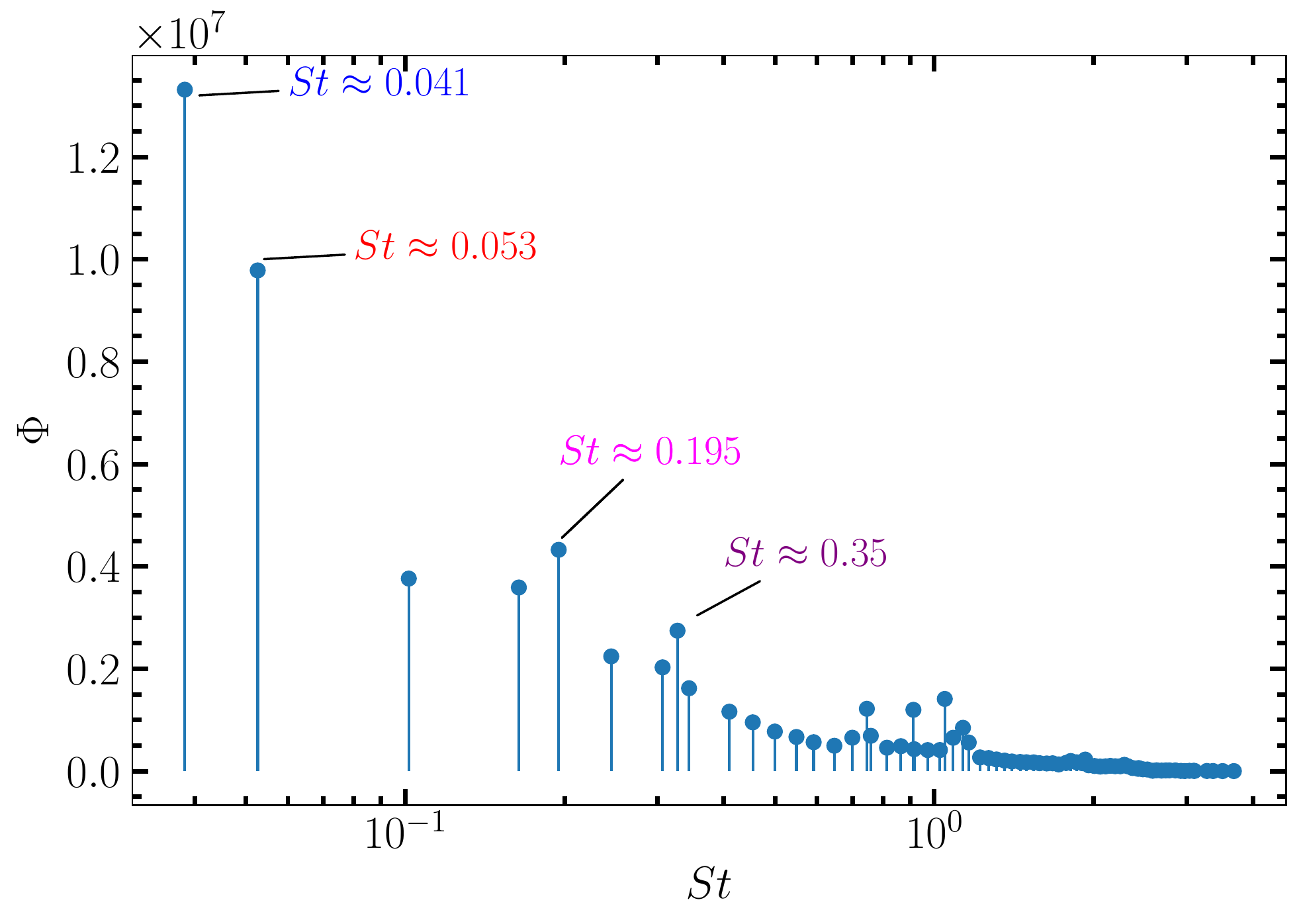}
         \caption{\label{fig:modesplot}}
         
     \end{subfigure}
        \caption{(a) DMD eigenvalue distribution on a unit circle in the complex plane. (b) Amplitude plot of the DMD modes against Strouhal number.   }
        \label{fig:three graphs}
\end{figure}

\begin{figure*}
     \centering
     \begin{subfigure}[b]{\columnwidth}
         \centering
         \includegraphics[width=\textwidth]{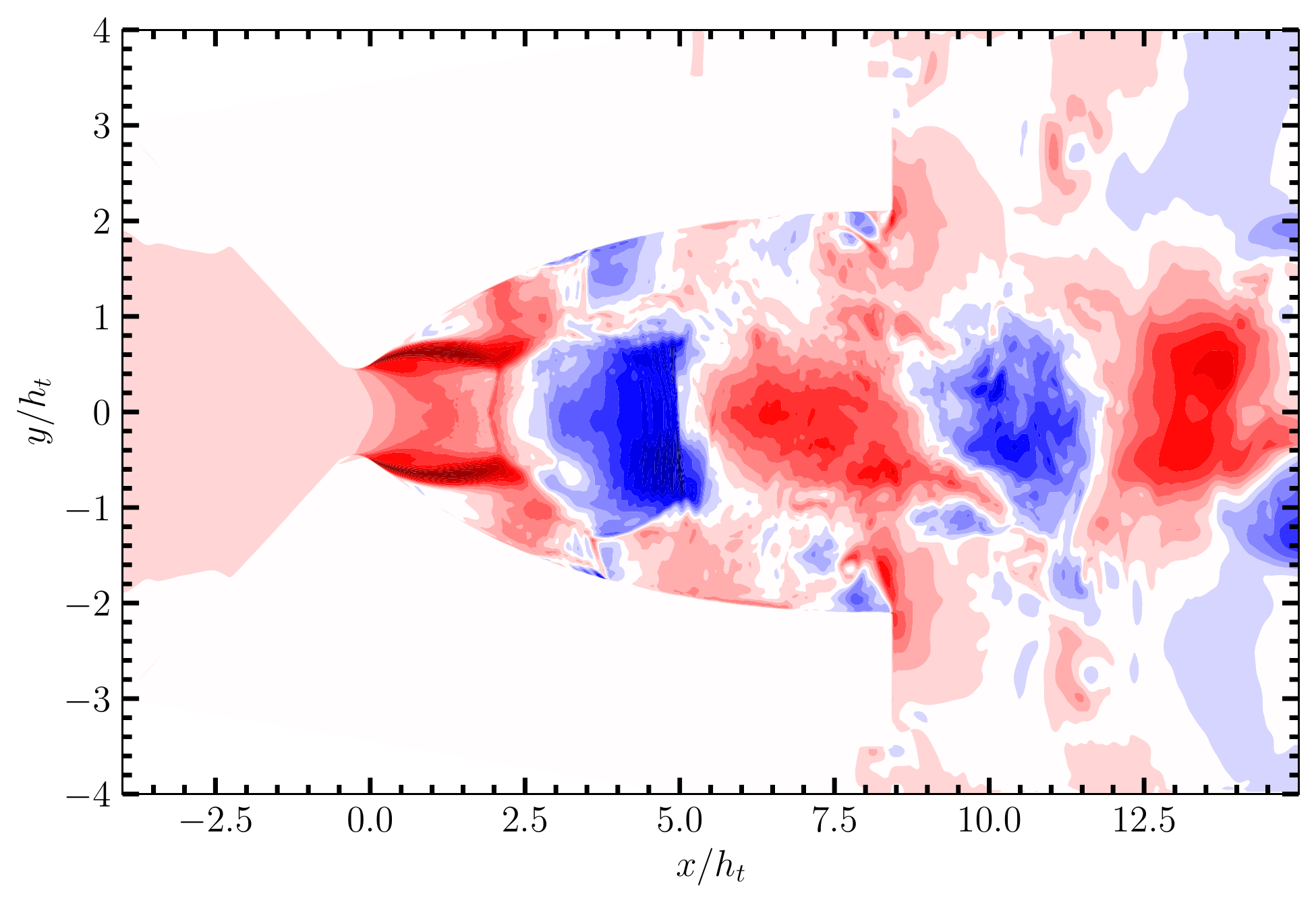}
         \caption{\label{fig:dmdeigenvalue}}
     \end{subfigure}
     \begin{subfigure}[b]{\columnwidth}
         \centering
         \includegraphics[width=\textwidth]{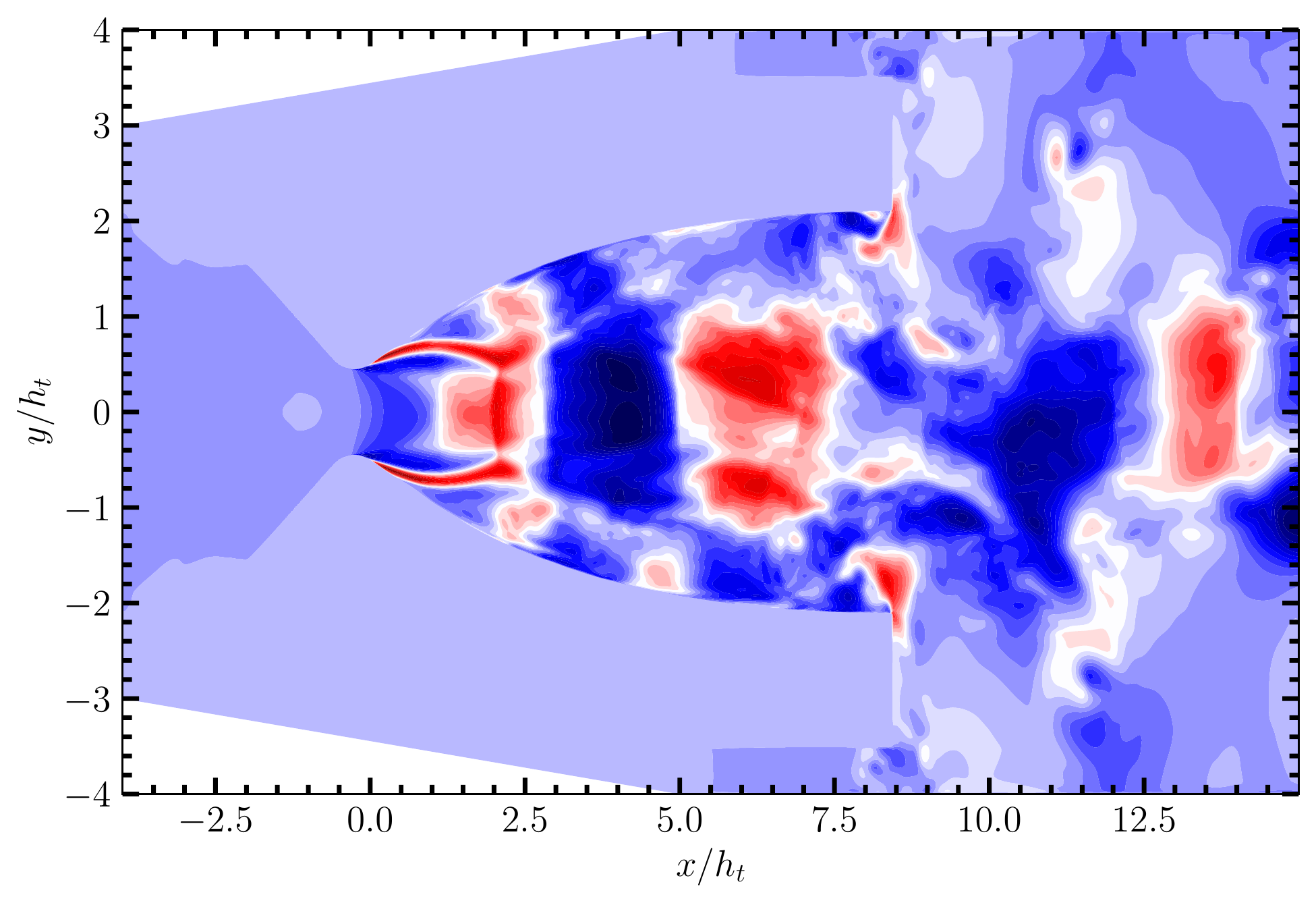}
         \caption{\label{fig:modesplot}}
         
     \end{subfigure}
          \centering
     \begin{subfigure}[b]{\columnwidth}
         \centering
         \includegraphics[width=\textwidth]{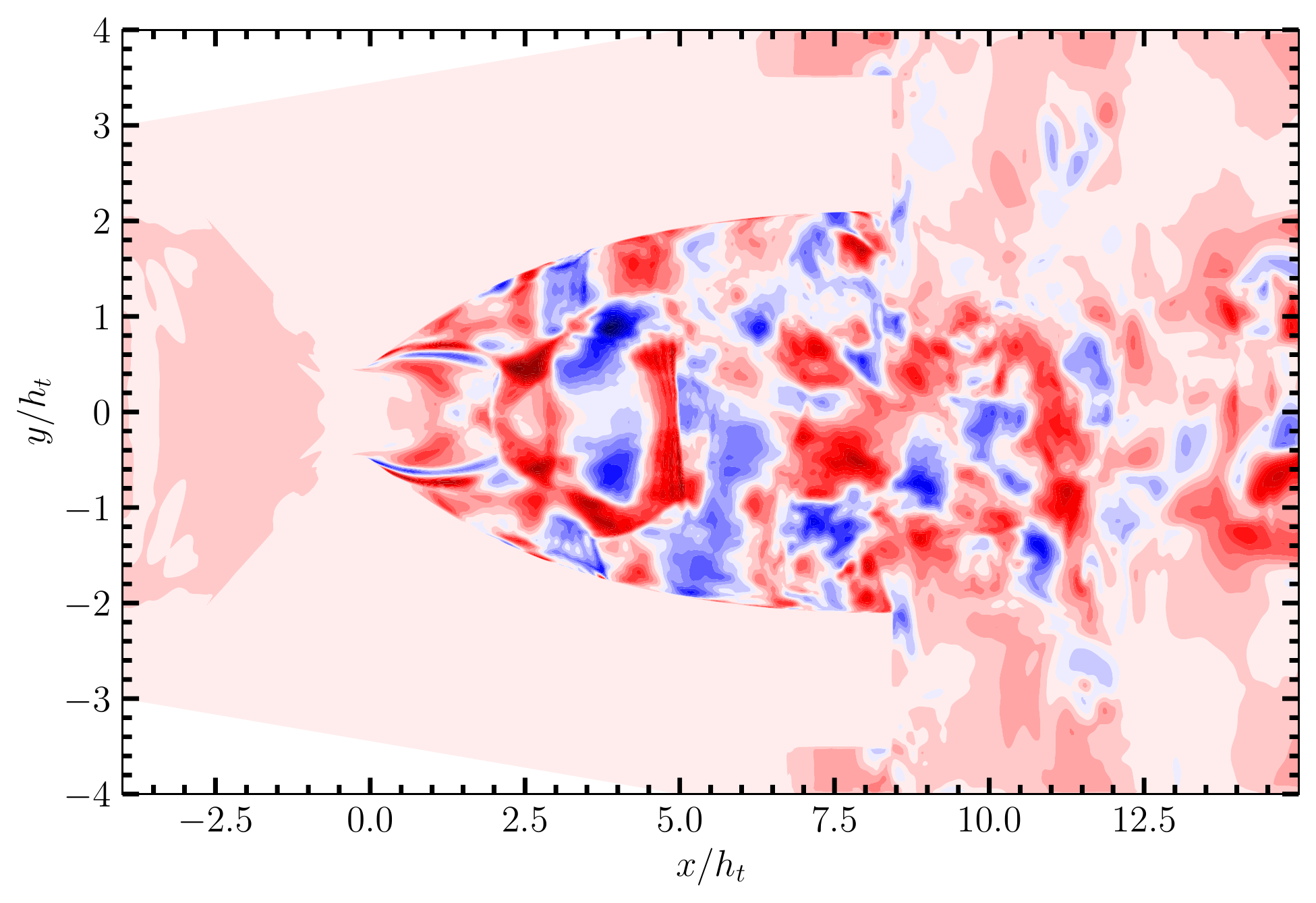}
         \caption{\label{fig:dmdeigenvalue}}
     \end{subfigure}
     \begin{subfigure}[b]{\columnwidth}
         \centering
         \includegraphics[width=\textwidth]{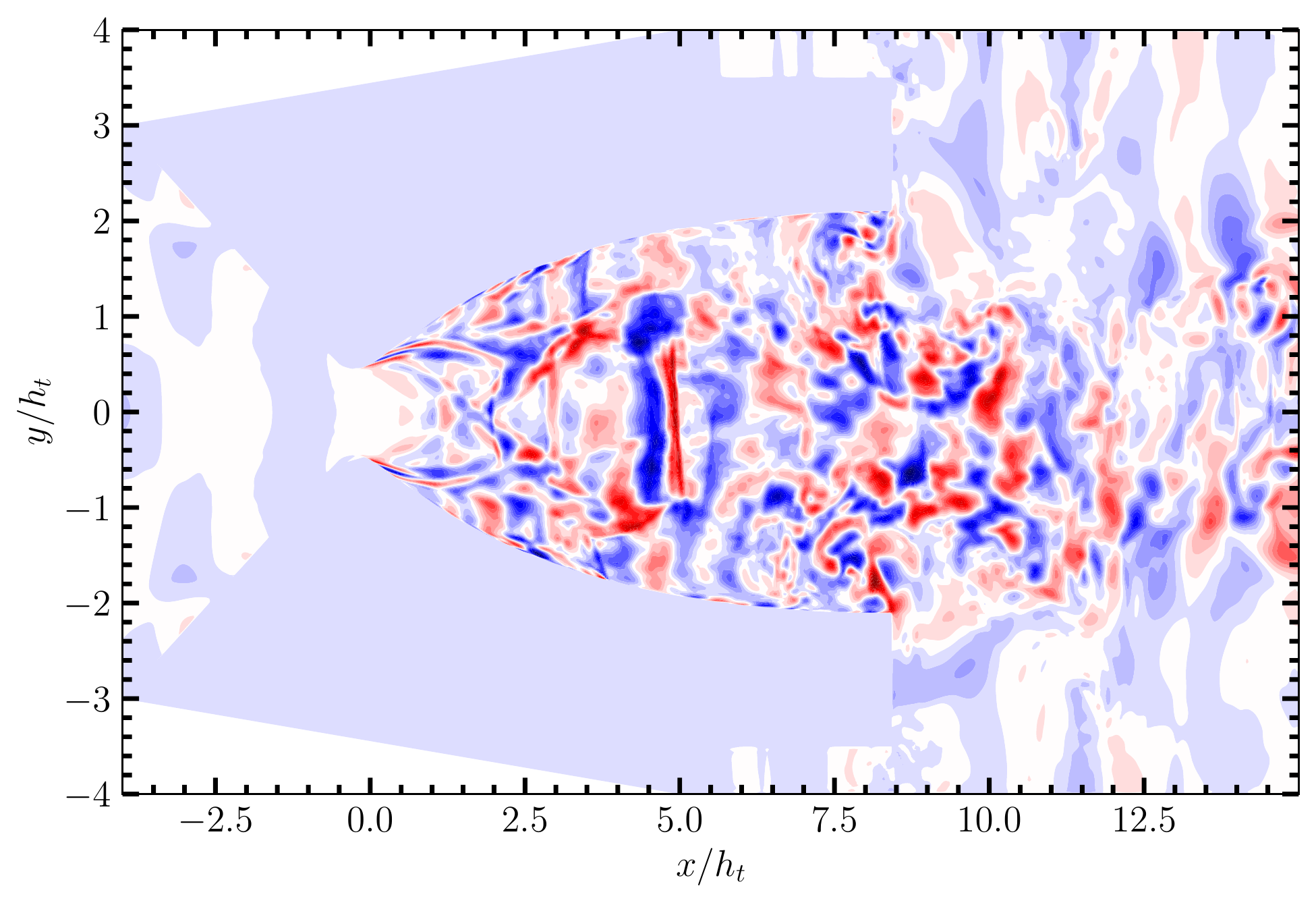}
         \caption{\label{fig:modesplot}}
         
     \end{subfigure}
        \caption{ Reconstructed DMD modes based on the velocity fluctuations (a) $St \approx 0.041 $ (b) $St \approx 0.053$ (c) $St \approx 0.195$ (d)~$ St\approx 0.35$, coloured by positive and negative fluctuations (red and blue, respectively)  }
        \label{fig:dmdmodes}
\end{figure*}

 \subsubsection{Model estimation of low-frequency unsteadiness via downstream mechanism}
 Before conducting further thorough analyses of the downstream mechanisms via reduced-order modal decomposition techniques, we further apply the simple model proposed by \citet{piponniau2009simple} for canonical SBLI configurations to our spectral measurements here. \citet{piponniau2009simple} suggested that the low-frequency unsteadiness is related to oscillatory shock motions, precipitating in large amplitude movements (i.e. expansion and contraction) of the downstream separation bubble as already discussed earlier. Therefore, the characteristic Strouhal number of the low-frequency movement must have functional dependence on the mixing-layer profile, which constantly ejects mass as a result of vortex shedding past the slip-layer or jet boundary (e.g. Kelvin-Helmholtz or related centrifugal instabilities), into the recirculation bubble; before it is once again entrained by the mixing layer to maintain an overall mean separated region.

 The model results in a characteristic Strouhal number, $St_{\ell}$ of the form:
 \begin{equation}
     St_{\ell} = \Phi(M_c) g(r,s)
 \end{equation}
 where $\Phi$ is the mixing layer spreading rate, which is a function of the convective Mach number, and $g(r,s)$ is given by:
 \begin{equation}
     g(r, s)=\frac{\delta_{r e f}^{\prime}}{2} \frac{(1-r)(1+\sqrt{s})}{(1+r \sqrt{s})}\left\{(1-r) C+\frac{r}{2}\right\}
 \end{equation}
 with $\delta_{r e f}^{\prime} \simeq 0.16 $, and where $r = u_2/u_1 $, $s = \rho_2/\rho_1$, with the subscripts $1$ and $2$ denoting the regions outside the separated shear layer, and the largest values obtained near the wall at incipient separation, respectively. Following the convection velocity estimate used by \citet{piponniau2009simple,gojon2016investigation,gojon2019antisymmetric}, we thus set $u_1 = u_j $ and $u_2 = u_e$ accordingly, with $C \simeq 0.14$ based on the error function similarity variable estimate of \citet{piponniau2009simple}, which agreed well with their experimental results. Also, the mixing layer spreading rate, $\Phi(M_c)$,  achieves a saturated value for convective Mach number values near unity \cite{smits2006turbulent}, which is not the case in the current simulations, even for $St \leq 0.06$ (see Fig.~\ref{fig:convecMc}). Thus, we use $\Phi(\bar{M}_c \simeq 0.6) \simeq 0.5$, based on the comprehensive experimental database available (cf. Fig. 2 in \citet{piponniau2009simple} which is adapted from Fig.~6.6 in \citet{smits2006turbulent} and references therein). This estimate yields $St_\ell \simeq 0.02 \pm 0.01$, which is again very close to that obtained from the transonic resonance model and the spectral data, providing further confidence that the low-frequency unsteadiness is supported by downstream mechanisms that serve to close the resonance loop.

\subsection{Dynamic Mode Decomposition}\label{sec:dmd}
Here we perform resolvent analysis using the reduced order modelling technique called Dynamic Mode Decomposition (DMD), which was developed by \citet{schmid2010dynamic}, and has been used widely in analysing coherent structures in fluid flows (see \citet{tu2013dynamic} for a comprehensive overview). This technique allows decomposition of the turbulent fluctuations in terms of its corresponding eigenfrequencies, which separates it from other techniques like the Proper Orthogonal Decomposition (POD)\citep{berkooz1993proper}, which ranks modes based on their energy content. Here we provide a brief reminder to introduce the relevant notation. These DMD modes used here are represented as a linear combination of $n$ spatial modes:
\begin{equation}
\boldsymbol{v'}(\boldsymbol{x}, t)=\sum_n b_n \mathrm{e}^{\lambda_n t} \boldsymbol{\phi'}_n(\boldsymbol{x})
\end{equation}
which have associated frequencies $\lambda_n$ and amplitudes $b_n$. We discuss the details of the algorithm and its significance below.

We use the Singular Value Decomposition (SVD) variant of the algorithm, in order to perform data-driven decomposition of the velocity fluctuations. The time-varying fluctuation $v'$, which are sampled at a fixed rate $\Delta t$ are first written into two snapshot data matrices of the form:
\begin{equation}\label{eqn:snapmatrix}
\begin{aligned}
V_0^{N-1}=\left[v_0, v_2, v_3, \ldots, v_{N-1}\right] \\
V_1^{N}=\left[v_1, v_2, v_3, \ldots, v_N\right]
\end{aligned}
\end{equation}
where $N$ stands for the total number of snapshots at one time instance, and $N-1$ corresponds to the number of snapshots at exactly one earlier time instance. The fundamental idea is that we assume the snapshot matrices satisfy a linear dynamical system such that:
\begin{equation}
    \mathbf{V}_{n+1}=\mathbf{A} \mathbf{V}_n
\end{equation}
where prior time snapshots are always related to future ones through a best-fit linear operator $\mathbf{A}$ in the least-square sense. In the continuous limit, this approximates the standard linear system solution:
\begin{equation}
v^{\prime}(t) = \exp({\mathbf{A}_{cont} t}) v_0
\end{equation}
Where $\mathbf{A}_{cont}$ is the continuous version of $\mathbf{A}$ and $v_0$  is an initial value. Thus, if the eigenvalues of $\mathbf{A}$ can be found, one can obtain the so-called DMD modes, which characterises the inherent frequencies of the flow \citep{rowley2017model}.

In practise, we do not compute $\mathbf{A}$ directly, but exploit the Moore-Penrose pseudoinverse:
\begin{equation}\label{eqn:eigendecomp}
    \mathbf{A} \triangleq \mathbf{V}^{\prime} \mathbf{V}^{*}
\end{equation}
and decompose the data matrix $V$ using the singular value decomposition (SVD)
\begin{equation}
V_1^{N-1} = U \Sigma W^*
\end{equation}
The quantity U is  a $M \times M$ unitary matrix, corresponding to the left singular vectors of the decomposition. These are also the POD modes. $\Sigma$ is the diagonal matrix of size $M \times N$, consisting of the singular values and $W^*$ is the unitary matrix of right singular vectors.
With the eigendecomposition of $\mathbf{A}$ as $\mathbf{A} = U\tilde{A}U^*$, the system can be reduced substantially, since a reduced version of linear operator $\tilde{A}$ is computed instead. This least-squared approximation minimises the so-called Frobenius norm, so that when substituting Eqn.\ref{eqn:eigendecomp} with its eigendecomposition:
\begin{equation}
\tilde{A} = U^*V^{\prime}W\Sigma^{-1}U^*U = U^*V^{\prime}W\Sigma^{-1}
\end{equation}
where $\tilde{A}$ is an $n \times n$ matrix based on the reduced SVD. In order to obtain the eigenvalues of $\mathbf{A}$, we solve the reduced eigenvalue problem:
\begin{equation}
    \tilde{A}v = v \Lambda
\end{equation}
According to \citet{tu2013dynamic}, the eigenvectors can be recovered by using:
\begin{equation}
\varphi = V^{\prime} W \Sigma ^ {-1} v 
\end{equation}
which can be simplified to 
\begin{equation}
\mathbf{A \varphi} = \varphi \Lambda 
\end{equation}
This allows us to obtain the eigenvalues of $\varphi$ (the DMD modes) in the matrix exponential. These eigenvalues (so-called Ritz eigenvalues) can be represented as:
\begin{equation}
    \exp({\lambda_i}) = \exp({\alpha_i + \beta_i j}) = \exp{\alpha_it} (\cos{\beta_it} + j\sin{\beta_i t} )
\end{equation}
One can observe if the real part (amplitude term) is non-zero, modes can be either growing or decaying depending on its sign, and if the imaginary part is non-zero, the eigenvalues will be time-periodic. We can also compute the frequencies of each eigenvalue directly from the sines and cosines. 

We compute the DMD modes using a serial Python code, with 400 velocity snapshots in total, while ensuring that the dominant frequencies predicted are convergent even with higher numbers of snapshots. Fig.~\ref{fig:dmdeigenvalue} displays the eigenvalues plotted on a unit circle in the imaginary plane. Almost all the values computed lie on the unit circle, which is expected for a well-saturated dynamical system (e.g., \citet{rowley2017model,taira2017modal}). The amplitude distribution (Fig.~\ref{fig:modesplot}) also indicates the dominant DMD modes within the flowfield, where the cumulative contribution of frequencies are concentrated at $St \approx 0.041$, $St \approx 0.053 $, $St \approx 0.195$ and $St \approx 0.35$. The lowest frequency tone is very close to that observed in the spectral analysis of $St \approx 0.0276$ within the detached shear layer, and shows that the transonic tone is well predicted via \citet{zaman2002investigation}'s relation. The other dominant tones, correspond quite closely to those found in the power spectra, where it is clear that the wall-pressure unsteadiness are also present within the global flowfield. Thus, we present the reconstructed DMD modes in Fig.~\ref{fig:dmdmodes}. First, we notice the clear periodicity in the perturbations observed for $St \approx 0.041 $, this is strongly representative of a standing wave mechanism, as shown in \citet{larusson2017dynamic}. Thus, the DMD predicts the standing pressure/velocity mode, and is able to reconstruct the behaviour associated with the transonic tone. The remaining plots, all confirm the dominant role of the separation bubble in the flow, where at all frequencies, signatures are always observed within it. It confirms that, even in the typical case where the Mach reflection does not occur at the axis (see e.g., the role of Mach reflection in the resonant loop in \citet{martelli2020flow} and \citet{bakulu2021jet}), the aeroacoustic feedback loop is now transferred to a separate location, and is supported in the separation bubble instead. These also feed the low frequency unsteadiness and importantly confirms that the downstream flow plays a significant role in sustaining the low frequency signatures, as well as the other dominant unsteadiness frequencies observed in the spectral measurements.

\subsection{Wall-pressure forces}
\begin{figure*}
     \centering
     \begin{subfigure}[b]{\columnwidth}
         \centering
         \includegraphics[width=\textwidth]{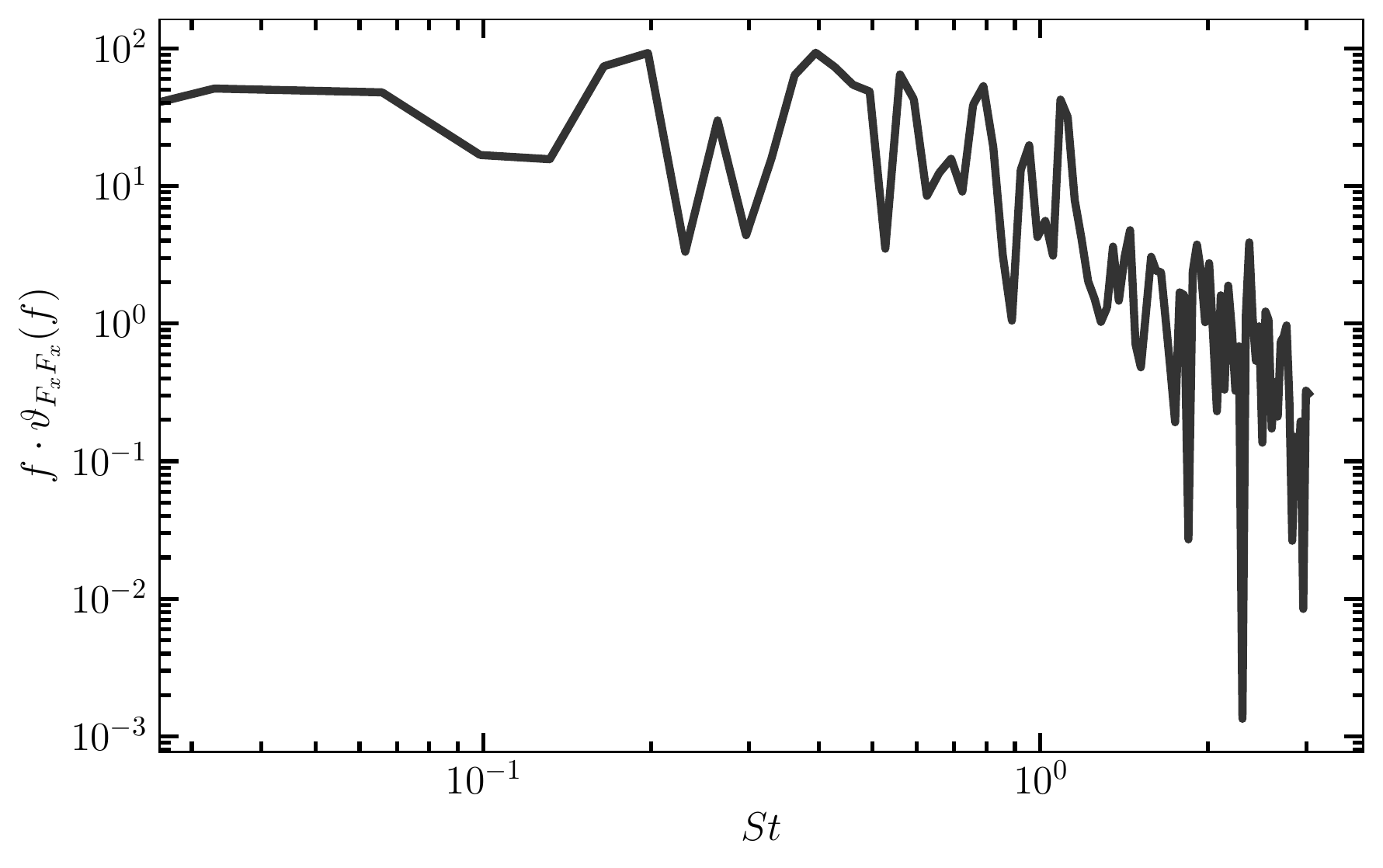}
         \caption{\label{fig:xforce}}
     \end{subfigure}
     \begin{subfigure}[b]{\columnwidth}
         \centering
         \includegraphics[width=\textwidth]{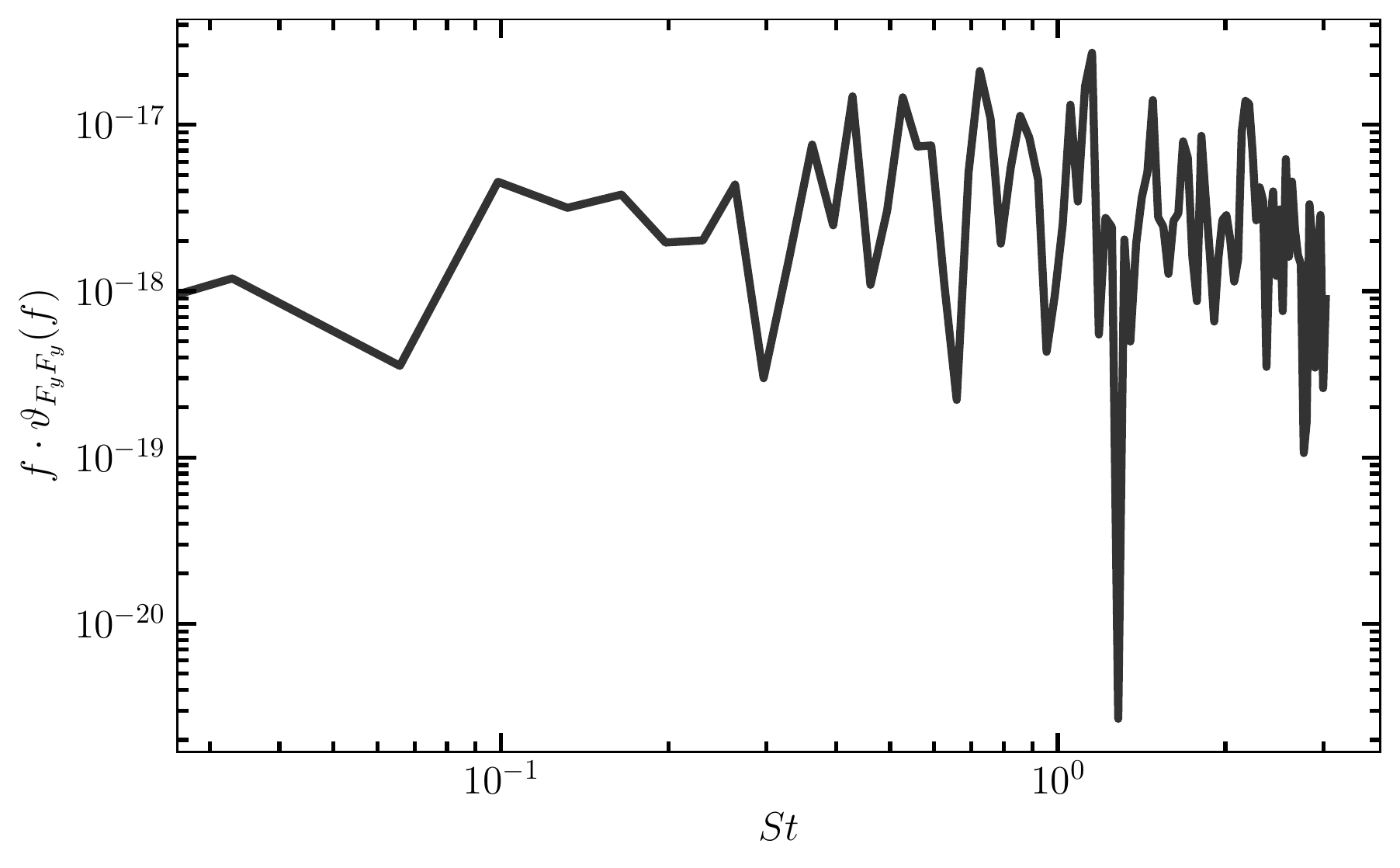}
         \caption{\label{fig:yforce}}        
     \end{subfigure}

 \caption{\label{fig:sideloads}(a) Pre-multiplied power spectra of the streamwise forces $f \cdot \vartheta_{F_xF-x} (f)$ (b) Same but for the wall-normal forces (side loads). }
\end{figure*}

Here we also briefly quantify the aerodynamic side loads within the flow, which are pronounced during flow separation since the wall pressure unsteadiness can majorly induce local lateral force generations. The computation of the streamwise and wall-normal forces are straightforward:
\begin{equation}
    \vec{F}=\int_c\left(p_w-p_{\infty}\right) \cdot \vec{n} \mathrm{~d} s
\end{equation}
and is a surface integral with a pressure differential corresponding to the pressure difference between the wall and the ambient medium. Fig.~\ref{fig:sideloads} displays the plot of the pre-multiplied power spectra for both the streamwise and lateral forces (side loads).It can be seen that $F_y$ contains comparable powers over all scales, consistent with the observations of \citet{zebiri2020shock}, thus suggesting no specific frequency contributes to the driving of aerodynamic side loads. This observation was also made from experiments by \citet{jaunet2017wall} for their TIC nozzle.  In light of the spectral analysis of the side loads, we here also propose to use the multi-resolution technique called continuous wavelet transform (CWT)  \citep{farge1992wavelet}, to perform joint time-frequency analysis. Here we use a complex Morlet wavelet, with scales varying from the minimum resolved frequency to the largest, with logarithmic intervals of base 2. Fig.~\ref{fig:sideloads} display the wavelet spectra for both components. The low frequency signature in the streamwise forcce component observed here is qualitatively similar to that of \citet{martelli2017detached} and \citet{martelli2019characterization}, where the resonant tone at the low Strouhal number persists throughout the time evolution, although recent work in canonical SBLis have suggested that more temporally intermittent structures can be observed (e.g., \citet{bernardini2022unsteadiness}). But it is not yet clear whether these features can also be resolved for planar nozzles. Nevertheless, the spectra seem to indicate some degree of temporal intermittency, which cannot be deduced from the Fourier spectra alone.
\begin{figure*}
     \centering
     \begin{subfigure}[b]{\columnwidth}
         \centering
         \includegraphics[width=\textwidth]{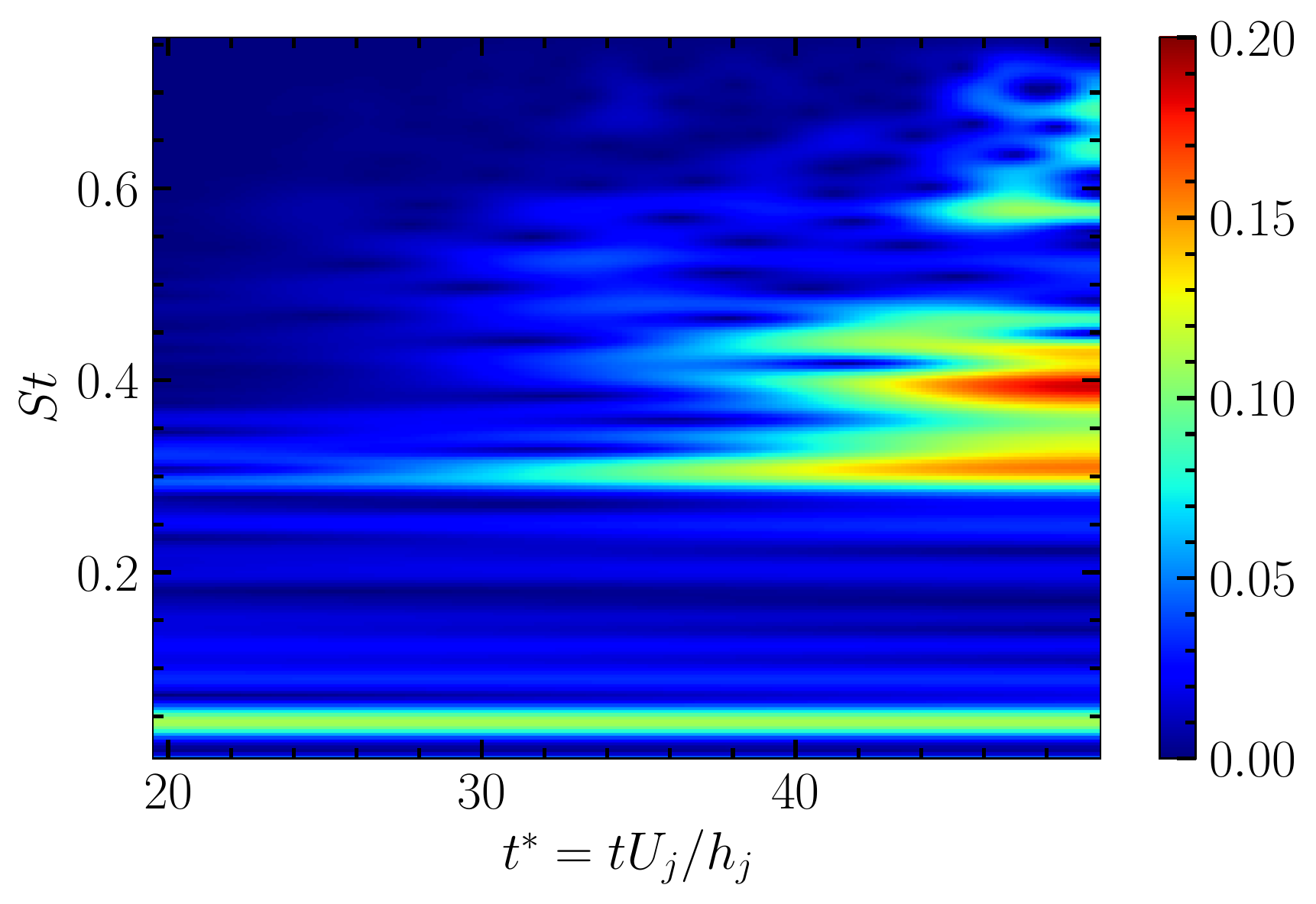}
         \caption{\label{fig:waveletx}}
     \end{subfigure}
     \begin{subfigure}[b]{\columnwidth}
         \centering
         \includegraphics[width=\textwidth]{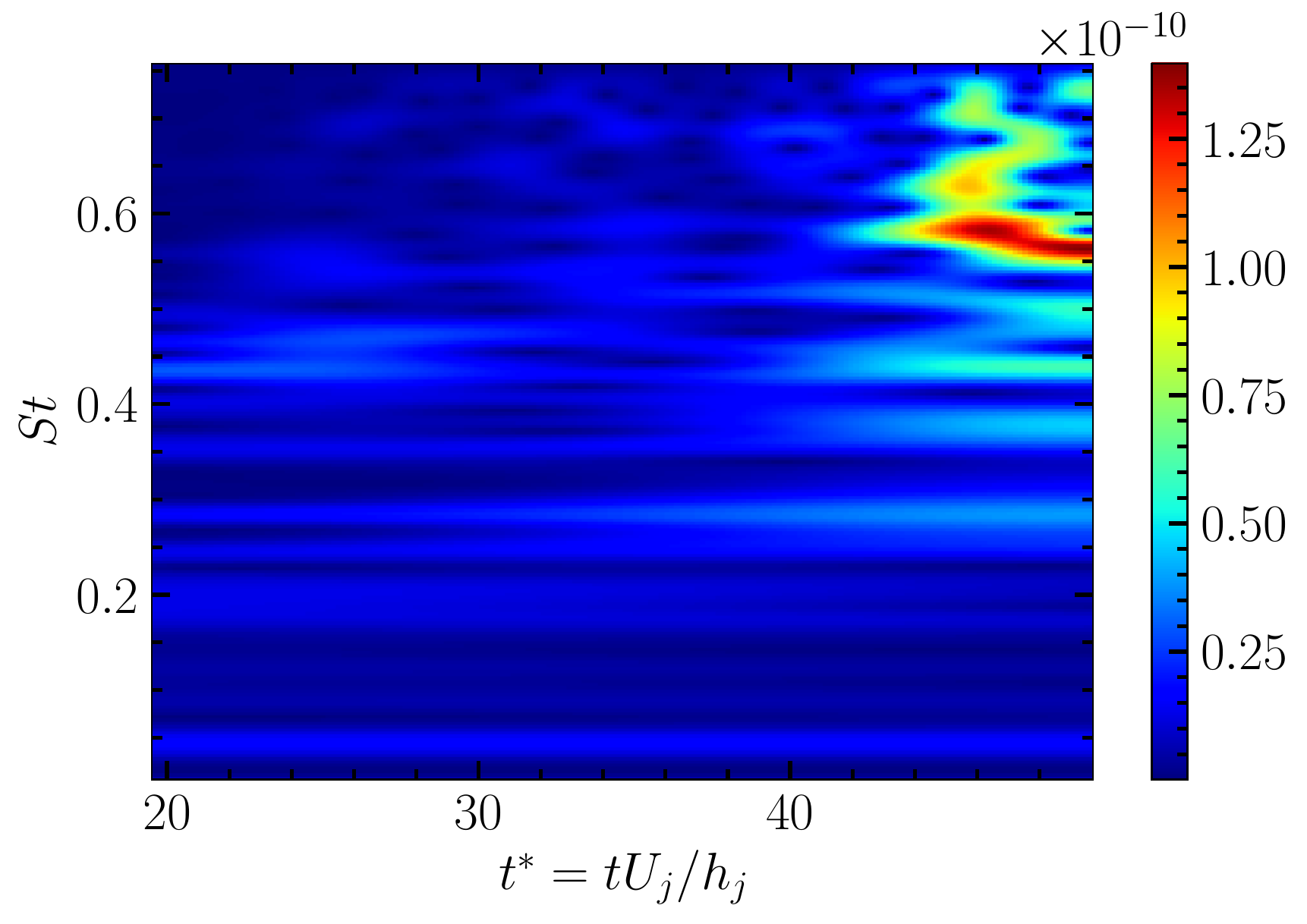}
         \caption{\label{fig:wavelety}}
         
     \end{subfigure}
        \caption{(a) Continuous wavelet transform scalogram for the streamwise force $F_{xx}$ (b) Same but for the wall-normal forces.  }
        \label{fig:three graphs}
\end{figure*}

\section{Conclusion}\label{sec:conclusion}
We performed a combined numerical and experimental study to investigate the spatiotemporal dynamics of an over-expanded transonic subscale nozzle with flow separation. The results obtained confirm a fundamental aeroacoustic feedback loop existing between the upstream separation location, and the downstream flow. The separation bubble not only plays a key role in sustaining this screech mechanism, but also continuously supports the existence of the low frequency unsteadiness through subsonic upstream and downstream propagating acoustic waves. Semi-empirical models and Dynamic Mode Decomposition (DMD) further supports this theory by confirming the existence of a resonance loop supported within it from pressure perturbations triggered by the reflected shock/turbulent-boundary-layer interactions. Such results demonstrate that regular reflection will only support upstream waves in the separation bubble, while Mach reflection has been hypothesised by \citet{martelli2020flow} to support propagation in two locations (the separation bubble and behind the Mach stem).  We also, for the first time, showed that it is possible for an over-expanded separated jet to support upstream propagating guided jet modes for closure of the screech loop through the vortex sheet analysis, which was previously only shown to be possible in unseparated jets\cite{chen2021flow}.

Wall-pressure unsteadiness analysis also affirm the existence of multiple characteristic frequencies within the flowfield, such as the low-frequency shock oscillation, vortex shedding frequency, shear layer unsteadiness frequency and others. While their effects are less pronounced in the transonic setting as compared to highly turbulent and supersonic SBLIs, they nonetheless serve to enhance and modulate wall-pressure perturbations, which eventually still result in the generation of lateral forces along the nozzle wall. These all contribute to unsteadiness in the global flowfield, and is never only constrained to the nozzle wall. 
\begin{acknowledgments}
The authors are grateful to Harald Kleine for assistance with the experiments and many helpful discussions regarding this work. We further thank Christine Charles, Horst Punzmann, Dimitrios Tsifakis, Thimthana Lee and Josef Richmond for useful discussions.  We also thank Juan Felipe Torres and Hua Xia for useful insights and conversations. J.K.J.H. acknowledges funding via the ANU Chancellor's International Scholarship and the Space Plasma, Astronomy and Astrophysics award. C.F.~acknowledges funding provided by the Australian Research Council (Future Fellowship FT180100495 and Discovery Projects DP230102280), and the Australia-Germany Joint Research Cooperation Scheme (UA-DAAD). We~further acknowledge high-performance computing resources provided by the Leibniz Rechenzentrum and the Gauss Centre for Supercomputing (grants~pr32lo, pr48pi and GCS Large-scale project~10391), the Australian National Computational Infrastructure in the framework of the National Computational Merit Allocation Scheme, ANU Merit Allocation Scheme (grant~ek9), and the ANU Startup Scheme (grant~xx52).

\end{acknowledgments}

\section*{Authors' Contributions }
\textbf{Justin Kin Jun Hew}: Conceptualization (lead); Data Curation (lead); Formal analysis (lead); Investigation (lead); Methodology (lead); Project administration (lead); Writing – original draft (lead); Software (lead); Visualization (lead); Writing - review and editing (equal). \textbf{Emanuele Martelli}: Writing – review and editing (equal); Validation (equal); Formal analysis (supporting). \textbf{Mahdi Davoodianidalik}: Data Curation (supporting); Project administration (supporting); Supervision (equal); Methodology (supporting). \textbf{Rod W. Boswell}: Supervision (equal); Funding acquisition (equal); Writing - review and editing (supporting). \textbf{Christoph Federrath}: Funding acquisition (equal); Writing - review and editing (supporting). \textbf{Matthew Shadwell} Methodology (supporting).
\section*{Data Availability Statement}
Simulation and experimental data produced in the present study can be made available upon reasonable request to the corresponding author. 
\appendix

\section{Q-criterion visualisation}\label{sec:qCriterion}

\begin{figure*}
     \centering
     \begin{subfigure}[b]{\textwidth}
         \centering
         \includegraphics[width=\textwidth]{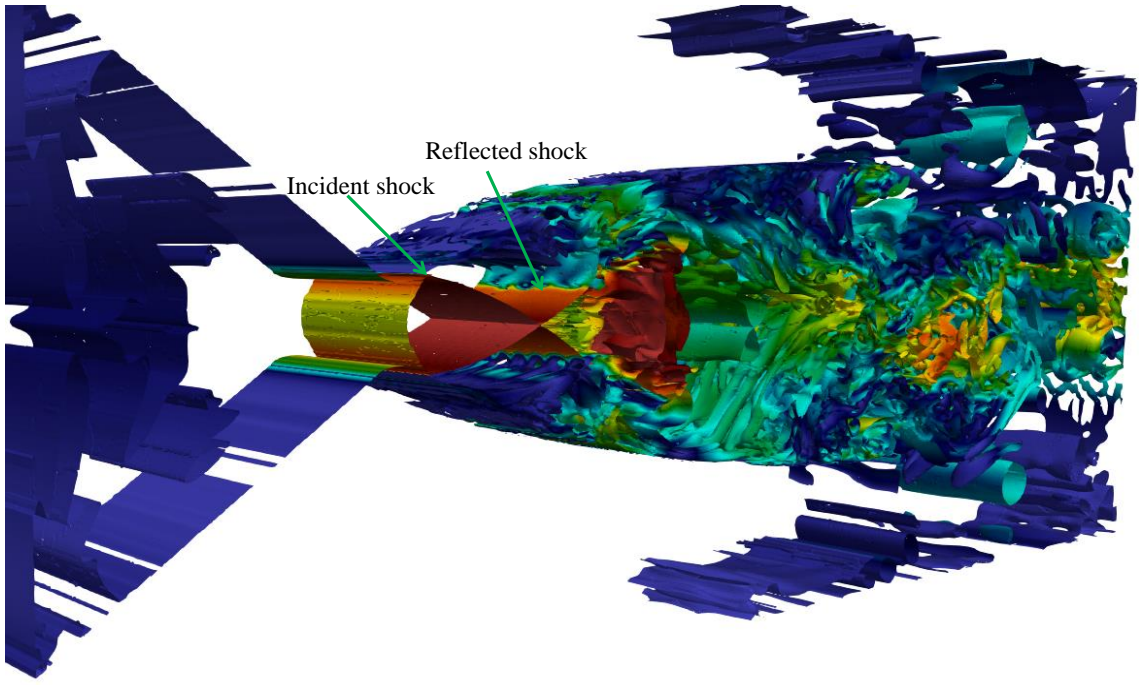}
         \caption{\label{fig:Bxprime}}
         
     \end{subfigure}
     \hfill
     \begin{subfigure}[b]{\textwidth}
         \centering
         \includegraphics[trim= 1cm 3cm 1cm 1cm,width=\textwidth]{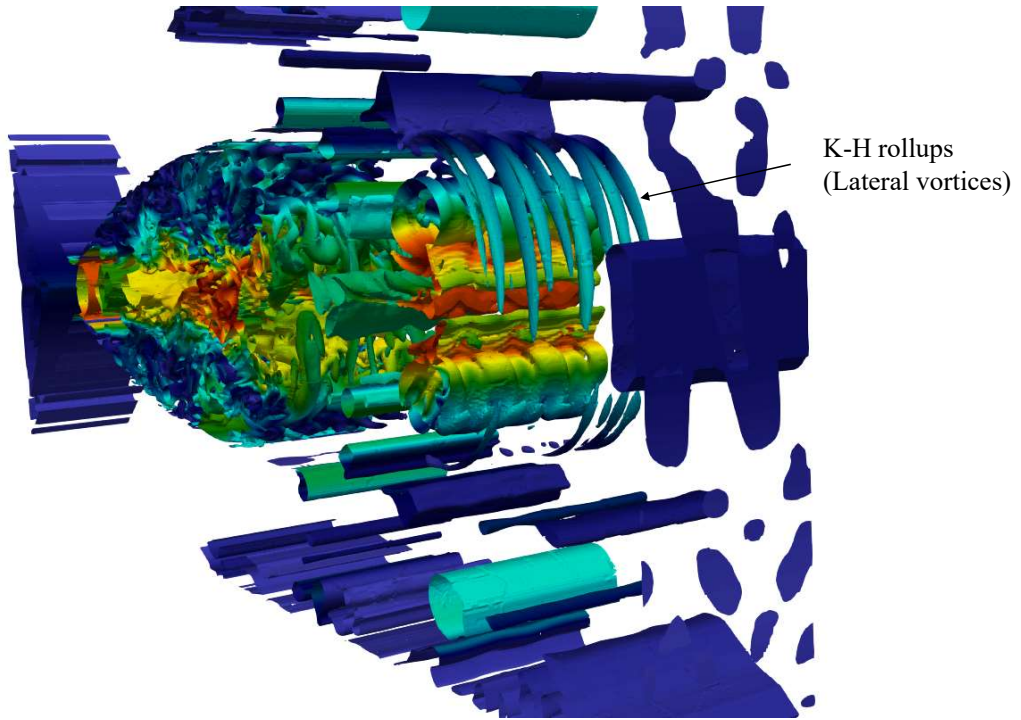}
         \caption{\label{fig:Byprime}}
         
     \end{subfigure}
        \caption{Iso-volume of the three-dimensional Q-criterion, overlaid with a numerical schlieren $\lvert \nabla \rho \rvert$, coloured by velocity magnitude $\lvert \mathbf{u} \rvert$.  }
        \label{fig:three graphs}
\end{figure*}

\bibliography{aipsamp}

\end{document}